%% file: main.tex
\documentclass[final]{siamart1116}

\input{shared}

\ifpdf
\hypersetup{
  pdftitle={\TheTitle},
  pdfauthor={\TheAuthors}
}
\fi

\begin{document}

\maketitle

\begin{abstract}
We present 2-D, 3-D, and spherical mesh generators for the Finite Element Method (FEM) using triangular and tetrahedral elements. The mesh nodes are treated as if they were linked by virtual springs that obey Hooke's law. Given the desired length for the springs, the FEM is used to solve for the optimal nodal positions for the static equilibrium of this spring system. A 'guide-mesh' approach allows the user to create embedded high resolution sub-regions within a coarser mesh. The method converges rapidly. For example, in 3-D, the algorithm is able to refine a specific region within an unstructured tetrahedral spherical shell so that the edge-length factor $l_{0r}/l_{0c} = 1/33$ within a few iterations, where $l_{0r}$ and $l_{0c}$ are the desired spring length for elements inside the refined and coarse regions respectively. One use for this type of mesh is to model regional problems as a fine region within a global mesh that has no fictitious boundaries, at only a small additional computational cost. The algorithm also includes routines to locally improve the quality of the mesh and to avoid badly shaped 'slivers-like' tetrahedra.
\end{abstract}



\section{Introduction}
\label{sec:Introduction}
Mesh generation and (adaptive) refinement are essential ingredients for computational modelling in various scientific and industrial fields. A particular design metric or goal is the quality of the generated mesh, because low-quality meshes can potentially lead to larger numerical approximation errors. A high-quality mesh would consist of triangles (in 2-D) or tetrahedra (in 3-D) that have aspect ratios near 1, i.e. their sides should have similar lengths. The techniques to generate meshes can be crudely classified into three groups: (1) The advancing front method \cite{Lohner1988, Schoberl1997,Choi2003,Ito2004} starts from the boundary of the domain. New elements are created one-by-one from an existing front of elements towards the interior until the region is filled. Advancing front methods generally create high-quality meshes close to the domain boundaries but can have difficulties in regions where advancing fronts merge. (2) Octree-based methods \cite{Mitchell1992,Labelle2007,Ito2009} produce graded meshes through recursive subdivision of the domain. The simplicity of these methods makes them very efficient. However, poorly shaped elements can be introduced near region boundaries. (3) Delaunay Triangulation ensures that the circumcircle/circumsphere associated to each triangle/tetrahedron does not contain any other point in its interior. This feature makes Delaunay-based methods \cite{Chew1989,Ruppert1995,Chew1997,Shewchuk1998} robust and efficient. However, in 3-D they can generate very poorly shaped tetrahedra with four almost coplanar vertex nodes. These so-called 'sliver' elements have a volume near zero. Several techniques to remove slivers have been proposed \cite{Cheng2000,Li2001,Cheng2002} although some slivers near the boundaries can typically persist \cite{Edelsbrunner2002}.

Current mesh generation algorithms oriented to engineering such as Gmsh \cite{Geuzaine2009}, GiD (\url{https://www.gidhome.com}) or TetGen \cite{Si2015} are based on the methods described above. Variational methods \cite{Alliez2005} rely on energy minimization to optimize the mesh during the generation procedure in order to create higher-quality meshes. A widely used open access community-code for 2-D mesh generation is Triangle \cite{shewchuk1996}, however there is no 3-D version of this mesh generator. DistMesh \cite{Persson2004} is an elegant and simple spring-based method that allows the user to create 2D and 3D unstructured meshes based on the distance from any point to the boundary of the domain. However this algorithm is often slow, requiring many steps to converge,

Any 'good' mesh should be able to meet the following requirements \cite{Bern1994}: (1) It conforms to the boundary; (2) It is fine enough in those regions where the problem to be solved demands higher accuracy; (3) Its total number of elements is as small as possible to reduce the size of the problem and the computational costs to solve it; (4) It has well-shaped elements to improve the performance of iterative methods such as the conjugate gradient method \cite{Shewchuk2002}. Frequently used mesh generators in 3-D geodynamic problems are the ones included in the ASPECT \cite{Kronbichler2012}, Rhea \cite{Burstedde2008} and Fluidity \cite{Davies2011} codes. ASPECT and Rhea are written in C++ with adaptive mesh refinement (AMR). However their regular hexahedral elements create so-called "hanging nodes" in regions where the resolution changes and cannot be directly applied to create well-formed tetrahedral elements. Fluidity is another example of AMR for a tetrahedral mesh. However it has very limited mesh generation capabilities, and in this context mesh-generation should not be confused with mesh adaptivity.

Here we present a new unstructured mesh generator that is based on a finite element implementation of the DistMesh approach using virtual springs between nodes and solving for the equilibrium positions of the nodes. We modify the Distmesh solution procedure to directly solve for static equilibrium. Our method is considerably faster than the DistMesh code. It also allows the user to create tetrahedral meshes without hanging nodes. The user can also create embedded high resolution sub-regions within a global coarse mesh. This approach becomes very useful when the goal is to create a mesh that minimizes the number of fictitious internal boundaries, within a computational problem.

A key design goal is the generation of a Delaunay mesh using a built-in MATLAB triangulation function called 'delaunay'. Throughout the algorithm, this 'delaunay' function is called to generate the spring connectivity matrix that relates nodes to triangles or tetrahedra. We have also developed and tested techniques for adding or rejecting nodes in regions where the mesh resolution is too high or too low respectively. A smooth variation in the element size between high resolution and low resolution regions is achieved by using a guide-mesh approach. These local operations improve the quality of the relatively few poorly shaped elements that can result from the ficticious spring algorithm to determine good nodal locations. The mesh-generation code is written in vectorized MATLAB, and can be easily used within the MATLAB working environment. 

We will present this approach first in its simplest form for making a mesh in a well-defined rectangular 2-D region (\Cref{sec:2-D Rectangular work flow}). In \Cref{sec:2-D Cylindrical annulus work flow} we show how a 2-D cylindrical annulus mesh can be generated with small modifications to the previous rectangular mesh generator algorithm. In \Cref{sec:3-D Spherical shell work flow} we present the modifications needed to create the 3-D spherical shell mesh that we are using to solve for mantle flow.

\section{2-D Rectangular work flow}
\label{sec:2-D Rectangular work flow}
\begin{figure}
	\centering
   	\includegraphics[width=0.7\textwidth]{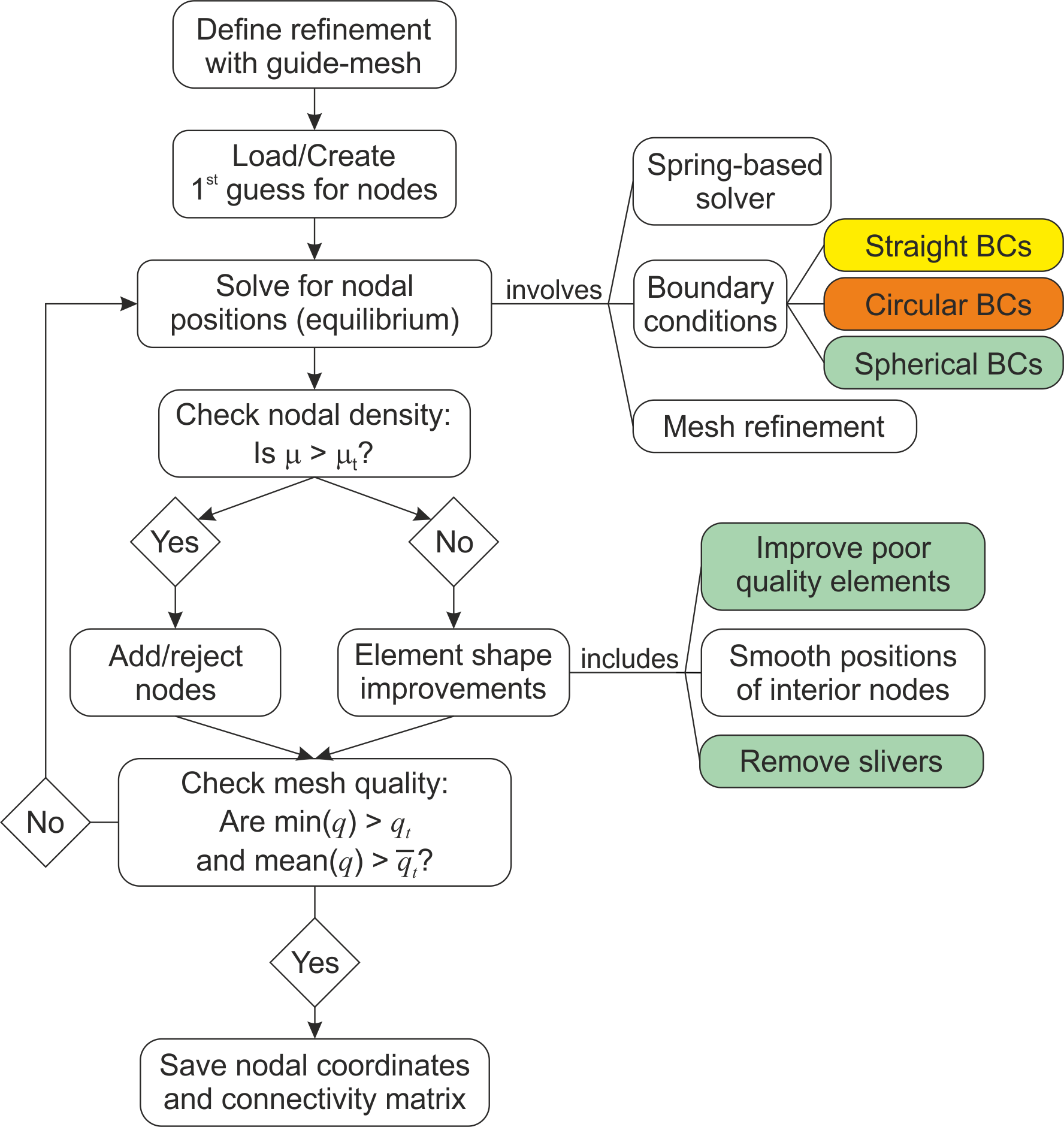}
  	\caption{Flow chart for the mesh generator iterative process. Yellow, orange and green boxes represent the routines exclusively used for creating 2-D rectangular meshes, 2-D cylindrical annulus meshes and 3-D spherical shell meshes, respectively. White boxes represent the shared routines to all mesh generators. $\mu$ is the mean of the misfit spring lengths (equation \cref{eq:16}) and $q$ is the quality factor of the elements (equations \cref{eq:13} and \cref{eq:31} for triangular and tetrahedral elements respectively). Tolerance parameters $\mu_t$, $q_t$ and $\bar{q}_t$ are listed in \cref{tab:1}.}
  	\label{fig:flow_chart}
\end{figure}

This mesh generation algorithm has its simplest form as a program to create a 2-D rectangular mesh with an embedded high resolution sub-region. The white and yellow boxes in \cref{fig:flow_chart} show the flowchart that describes this algorithm.

\subsection*{Step 1: Definition of preferred nodal distances and initial placement of the nodes}
\label{sec:Definition of preferred nodal distances and initial placement of the nodes}
\begin{figure}
    \centering
    \includegraphics[width=1\textwidth]{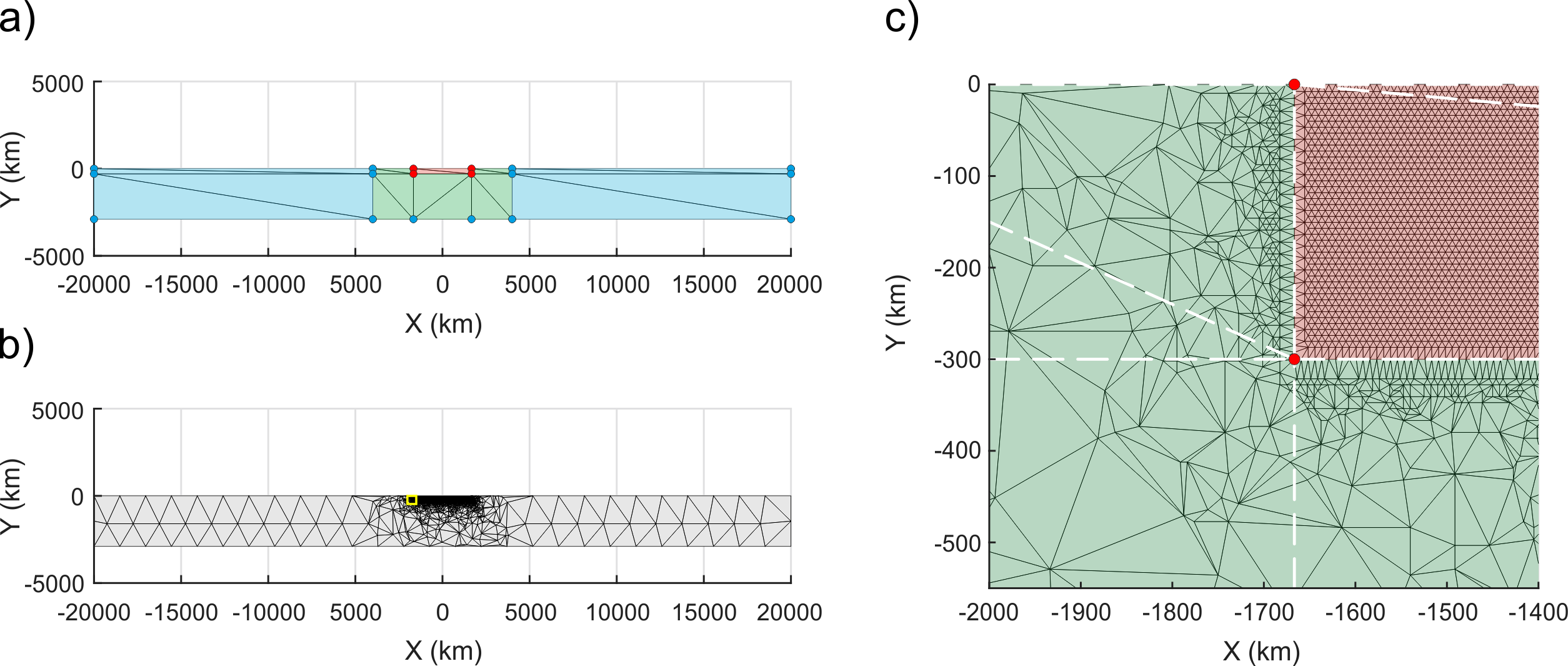}
    \caption{(a) Guide-mesh defined by a few nodes in Cartesian coordinates for a rectangular mesh. The parameters for this mesh are listed in \cref{tab:1}. Each node is assigned a value for the desired spring length, being $l_{0r}$ for red dots and $l_{0c}$ for blue dots. The length of the springs within the refined region (in red) is approximately equal to $l_{0r}$. The length of the springs within the transition region (in green) varies smoothly from $l_{0r}$ to $l_{0c}$. The length of springs within the coarse region (in blue) is approximately equal to $l_{0c}$. (b) Initial guess for the rectangular mesh. (c) Zoom around the left boundary of the refined region for the initial guess (yellow line in (b)). The guide-mesh defining refined (red) and transition (green) regions is shown in white dashed lines.}
    \label{fig:RECT_guide_mesh_and_first_guess}
\end{figure}

The first step in this recipe is to define the preferred nodal distances within the refined ($l_{0r}$) and  coarse ($l_{0c}$) regions as well as the dimensions of the regions. In order to avoid poor quality elements, an appropriate smooth transition for the mesh refinement should be specified. Here we choose a preferred spring-length function that is defined on a so-called 'guide-mesh'. This approach is very similar to the background grid approach created by \cite{Lohner1988}. The generation of a refined rectangular mesh using the guide-mesh approach involves the following steps. First, create a (coarse) mesh to serve as a guide-mesh with only a small number of nodes defining the boundaries of the domain and the internal boundaries of the embedded high resolution and transition sub-regions. Second, create the design function $l_0(x$, $y)$ for each node of the guide-mesh. This function defines the desired length for the springs around those points. Third, the function $l_0(x$, $y)$ is evaluated at the midpoint of all springs using linear Finite Element shape functions. We find that a coarse guide-mesh is a simple and flexible way to control nodal spacing during the generation of a Finite Element mesh. \cref{fig:RECT_guide_mesh_and_first_guess}a shows the guide-mesh for a rectangular mesh example whose parameters are listed in \cref{tab:1}. Red and blue dots represent nodes in the guide-mesh with defined $l_{0r}$ and $l_{0c}$, respectively. The red region represents the refined region of the mesh with spring length approximately equal to $l_{0r}$. The green region defines the transition region where the length of the springs smoothly varies from $l_{0r}$ to $l_{0c}$. The blue region represents the coarse region of the mesh with a apporximate spring length of $l_{0c}$.

The next step is to create a starting guess for the locations of the nodes. Computational work is reduced considerably with a good initial guess for the density of the nodes. Nodes on the boundary and within the domain are created taking into account both the location of the refined region and the desired springs length for elements inside the refined and coarse regions. Boundary nodes in the refined and coarse regions are created using $l_{0r}$ and $l_{0c}$ respectively for the spacing between the nodes. The interior nodes within the refined and coarse regions are created using a circle packing lattice with radius equal to $l_{0r}/2$ and $l_{0c}/2$ respectively. This fills each region with an equilateral triangular tiling. In the transition region the size of the elements is expected to change smoothly between $l_{0r}$ and $l_{0c}$. The initial placement for boundary and interior nodes in the transition region is created using $l_{0r}$ as explained above. After this step, the rejection method described in \cite{Persson2004} is used to discard points and create a 'balanced' intitial distribution of nodes. After performing a Delaunay triangulation, a quasi-regular mesh of triangles within the refined and coarse regions, with a poorly structured transition region between them is created (\cref{fig:RECT_guide_mesh_and_first_guess}b). \cref{fig:RECT_guide_mesh_and_first_guess}c shows a zoom of the initial mesh with the guide-mesh also shown.

\subsection*{Step 2: Spring-based solver}
\label{sec:Spring-based solver}
\begin{figure}
    \centering
    \includegraphics[width=1\textwidth]{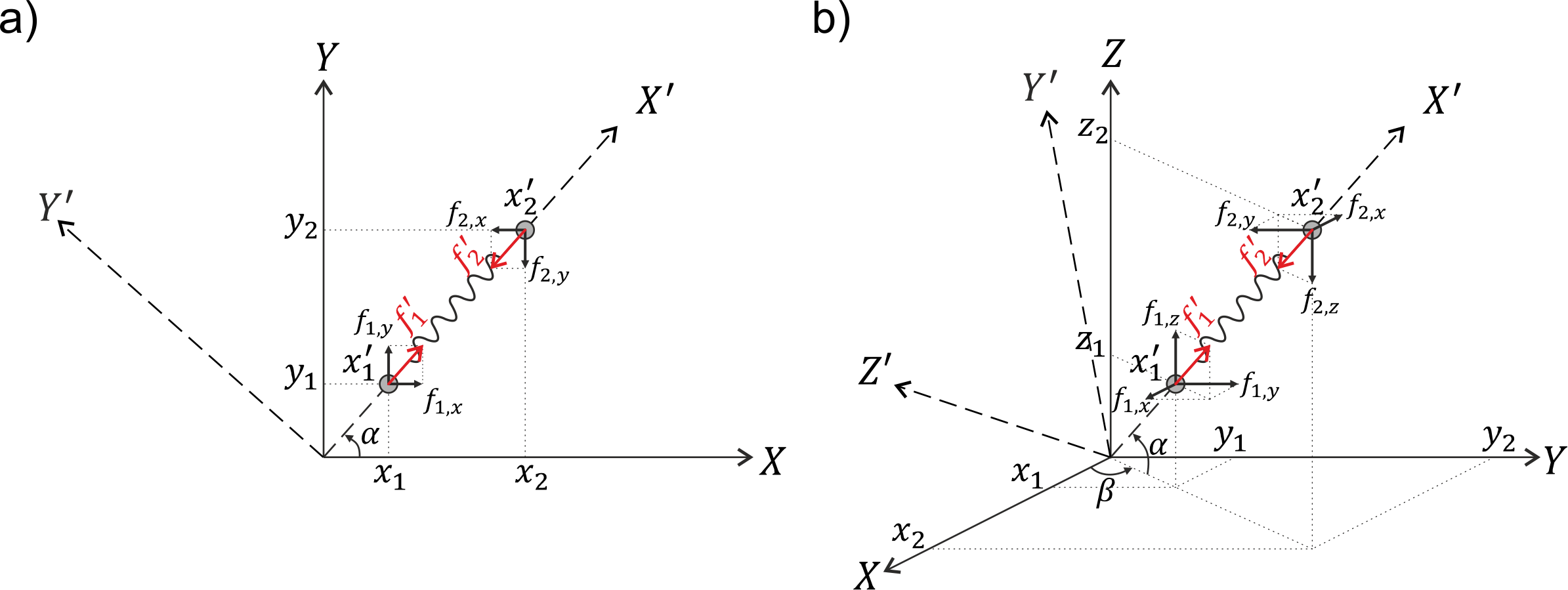}
    \caption{(a) Virtual spring in the 2-D space. Both global reference system ($X$, $Y$) and local reference system ($X'$, $Y'$) are shown. (b) Virtual spring in the 3-D space. Both global reference system $(X$, $Y$, $Z)$ and local reference system $(X'$, $Y'$, $Z')$ are shown. Grey dots represent two nodes linked by the virtual spring. Red arrows represent the forces acting at each end of the spring.}
    \label{fig:springs_2D_and_3D}
\end{figure}

Inspired by \cite{Persson2004}, to generate an unstructured mesh we link the future locations of finite element nodes with virtual elastic springs. The spring length is used to define the desired nodal distance within any mesh region, i.e., short springs lead to mesh regions with higher resolution and longer springs lead to lower resolution mesh regions. Nodal positions are solved for so that the global network of virtual springs is in static equilibrium. The behaviour of each ficticious spring is described by Hooke's law 
\begin{equation}
  	\label{eq:1}
    F=-k\delta s
\end{equation}
where $F$ is the force acting at each end of spring, $k$ is the stiffness of the spring, and $\delta s$ is the distance the spring is stretched or compressed from its equilibrium length $l_0$. Forces and nodal positions are expressed in $x$, $y$ coordinates in 2-D (\cref{fig:springs_2D_and_3D}a). Because Hooke's law is formulated along the spring direction it is necessary to introduce the $X'$ axis as the local 1-D reference system to solve for the nodal positions. Hooke's law for each spring in the local 1-D reference system is given by
\begin{subequations}
	\begin{align}
		\label{eq:2a}
        	{f_1}' &= \ \ \,k \delta s = \ \ \,k({x_2}'-{x_1}'-l_0) \\
        \label{eq:2b}
        	{f_2}' &= -k \delta s = -k({x_2}'-{x_1}'-l_0)
	\end{align}
\end{subequations}
where $f'$ and $x'$ are the force and position of the ends of the spring given by the subscripts 1 and 2, respectively. 
Writing equations \cref{eq:2a,eq:2b} in matrix form, and moving the force terms to the left hand side yields
\begin{equation}
  	\label{eq:3}
    \left(\begin{array}{c}
    {f_1}' \\
	{f_2}'
	\end{array}\right)
    + k
	\left[\begin{array}{rr}
		-1 & 1\\
		1 & -1
	\end{array}\right]
	\left(\begin{array}{c}
		0\\
		l_0
	\end{array}\right)
	= k
	\left[\begin{array}{rr}
		-1 &  1 \\
		 1 & -1
	\end{array}\right]
	\left(\begin{array}{c}
		{x_1}' \\
		{x_2}'
	\end{array}\right)
\end{equation}
In order to solve for the nodal positions in 2-D, a change from local coordinates (${x_1}'$, $0$; ${x_2}'$, $0$) to global coordinates ($x_1$, $y_1$; $x_2$, $y_2$) is needed. This change of coordinates is described in matrix form as
\begin{equation}
  	\label{eq:4}
  	\boldsymbol{R_{2D}} =
  	\left[\begin{array}{cccc}
  	\cos\alpha & \sin\alpha &          0 &          0 \\
  	         0 &          0 & \cos\alpha & \sin\alpha
  	\end{array}\right]
\end{equation}
where $\alpha$ is the angle of the $X'$ axis measured from the $X$ axis in the counterclockwise direction (\cref{fig:springs_2D_and_3D}a). Applying equation \cref{eq:4} to equation \cref{eq:3} (see \cref{SM1} for further details), equation \cref{eq:3} becomes
\begin{equation}
	\label{eq:5}
	k	
	\left[\begin{array}{cccc}
		 -{c_\alpha}^2       &  -s_\alpha c_\alpha &  {c_\alpha}^2 	     &   s_\alpha c_\alpha  \\
          -s_\alpha c_\alpha & -{s_\alpha}^2  	   &   s_\alpha c_\alpha &  {s_\alpha}^2  		\\
          {c_\alpha}^2       &   s_\alpha c_\alpha & -{c_\alpha}^2       &  -s_\alpha c_\alpha  \\
           s_\alpha c_\alpha &  {s_\alpha}^2  	   &  -s_\alpha c_\alpha & -{s_\alpha}^2		 
	\end{array}\right] 
	\left(\begin{array}{c}
		x_1 \\
		y_1 \\
		z_1 \\
		x_2 
	\end{array}\right)
	=
	\left(\begin{array}{c}
		f_{1,x} \\
		f_{1,y} \\
		f_{2,x} \\
		f_{2,y} 
	\end{array}\right)
	+
	kl_0
	\left(\begin{array}{c}
		 c_\alpha \\
		 s_\alpha \\
		-c_\alpha \\
		-s_\alpha
	\end{array}\right)
\end{equation}
where $s_\alpha\equiv\sin\alpha$ and $c_\alpha\equiv\cos\alpha$. Equation \cref{eq:5} can be written in the matrix form as
\begin{equation}
  	\label{eq:6}
  	\boldsymbol{K} \boldsymbol{x} = \boldsymbol{f} + \boldsymbol{f_{l_0}}
\end{equation}
where $\boldsymbol{K}$ is the stiffness matrix, $\boldsymbol{x}$ is the nodal displacement vector, $\boldsymbol{f}$ is the element force vector and $\boldsymbol{f_{l_0}}$ is the force-term created by the fact that the springs would have zero-force at their desired length. Because the system of equations is solved for its equilibrium steady state, $\boldsymbol{f} = 0$. A vectorized 'blocking' technique based on the MATLAB methodology described in the MILAMIN code \cite{Dabrowski2008} is employed to speed up the assembly of the stiffness matrix. The solution to this problem is the 'optimal' position of each node obtained from the inversion of the system of static force equilibrium equations
\begin{equation}
  	\label{eq:7}
    \boldsymbol{x} = \boldsymbol{K}^{-1} \boldsymbol{f_{l_0}}
\end{equation}

\textit{\textbf{Straight line Boundary Conditions}}. Boundary conditions are necessary to constrain the mesh to the desired domain boundaries, and to differentiate between boundary and interior nodes. In the simple case of a rectangular mesh, a boundary node is free to slide along a domain edges parallel to the $X$- or $Y$-axis. We achieve this by setting one of its $y_i$ or $x_i$ values to be fixed and letting the other value vary so that the node is free to move along the boundary segment. In the case of a general line that is not parallel to the $X$- or $Y$-axes, this requires a transformation from global coordinates to a new local coordinate system in which the constraint direction is parallel to a local coordinate axis. In other words, the new local axes have to be parallel to and perpendicular to the boundary segment. 
\begin{figure}
  	\centering
  	\includegraphics[width=0.4\textwidth]{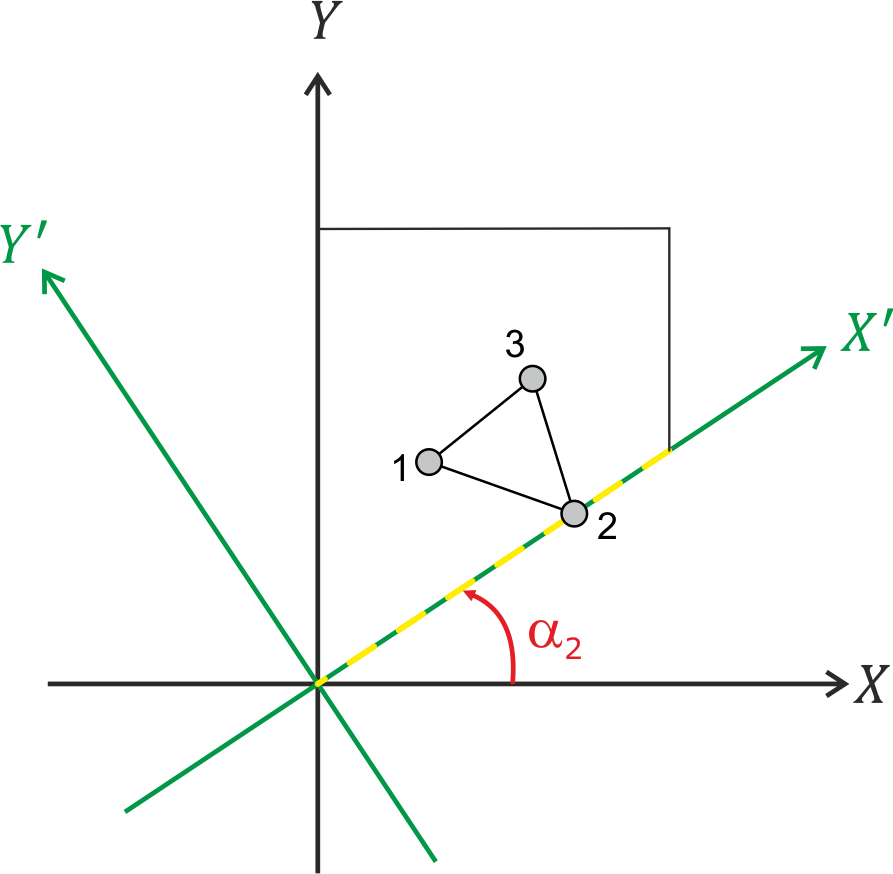}
  	\caption{Implementation of boundary conditions along a straight tilted segment (yellow dashed line) for one triangle. A rotation is needed for the node 2 in order to pass from the global reference system ($X$, $Y$) to the local reference system ($X'$, $Y'$) where $ {y_2}'= 0 $ is the constrained boundary condition.}
    \label{fig:BCs_straight_sloped_line}
\end{figure}
For simplicity, the mathematical implementation is shown for one triangle (\cref{fig:BCs_straight_sloped_line}). Node 2 is free to slide along the tilted segment (yellow dashed line in \cref{fig:BCs_straight_sloped_line}) since $ {y_2}'= 0 $ defines the boundary constraint. The boundary condition is imposed by a rotation of coordinate system for node 2 given by the transformation matrix $ \boldsymbol{T} $ that relates global coordinates $ \boldsymbol{x} $ to local coordinates $ \boldsymbol{x}' $ by
\begin{equation}
\label{eq:8}
	\underbrace{\left(\begin{array}{c}
		x_1 \\
		y_1 \\
		x_2 \\
		y_2 \\
		x_3 \\
		y_3
	\end{array}\right)}_{\boldsymbol{x}}
	=
	\underbrace{\left[\begin{array}{rrrrrr}
	1\  								       \\
	\ & 1\  							       \\
    \ & \ & \cos{\alpha_2} & -\sin{\alpha_2} \ \\
	\ & \ & \sin{\alpha_2} &  \cos{\alpha_2} \ \\
	\ & \ & \ & \ & 1\ \   			           \\
	\ & \ & \ & \ & \ & 1\ \
	\end{array}\right]}_{\boldsymbol{T}}
	\underbrace{\left(\begin{array}{c}
		x_1    \\
		y_1    \\
		{x_2}' \\
        0      \\
		x_3    \\
		y_3
	\end{array}\right)}_{\boldsymbol{x}'}
\end{equation}

Applying the transformation matrix to the stiffness matrix and force vector
\begin{equation}
  	\label{eq:9}
    \boldsymbol{K}' = \boldsymbol{T}^{\mkern-1.5mu\mathsf{T}} \boldsymbol{K} \boldsymbol{T}
\end{equation}
\begin{equation}
    \label{eq:10}
    \boldsymbol{f_{l_0}}' = \boldsymbol{T}^{\mkern-1.5mu\mathsf{T}} \boldsymbol{f_{l_0}}
\end{equation}
the new system of equations is given by
\begin{equation}
    \label{eq:11}
    \boldsymbol{K}' \boldsymbol{x}' = \boldsymbol{f_{l_0}}'
\end{equation}
which is solved for $ \boldsymbol{x}' $ . When desired, the original global coordinates are recovered through the transformation matrix
\begin{equation}
    \label{eq:12}
    \boldsymbol{x} = \boldsymbol{T} \boldsymbol{x}'
\end{equation}

\subsection*{Step 3: Mesh refinement}
\label{Mesh refinement}
In this algorithm we refine a mesh by decreasing the element size in the region of interest. One common issue in the refinement process arises from the size contrast between large and small elements within a short spatial interval so that poorly-shaped elements with short and long edges may form. In order to mitigate this issue a transition region surrounding the refined region is defined using the guide-mesh approach described above (see \cref{fig:RECT_guide_mesh_and_first_guess}a).

\textit{\textbf{Quality factor for triangles}}. The 'quality' of a mesh is determined by assessing the quality of its individual elements. This usually involves measures of angles, edge lengths, areas (in 2-D), volumes (in 3-D), or the radius of its inscribed and circumscribed circles/spheres, see e.g., \cite{Dompierre1998, Shewchuk2002}. Here we use a normalized quality factor, which in 2-D is given by
\begin{equation}
    \label{eq:13}
    q_{_{2D}} = \frac{2r_c}{R_c}
\end{equation}
where $ r_c $ is the radius of the element's inscribed circle and $ R_c $ is the radius of its circumscribed circle. $ R_c $ and $ r_c $ can be expressed as
\begin{equation}
    \label{eq:14}
    r_c
    =
    \frac{1}{2}
    \sqrt[]{
    \frac{(b+c-a)(c+a-b)(a+b-c)}{a+b+c}}
\end{equation}
\begin{equation}
    \label{eq:15}
    R_c
    =
    \frac{abc}{\sqrt[]{(a+b+c)(b+c-a)(c+a-b)(a+b-c)}}
\end{equation}
where $a$, $b$ and $c$ are the side lengths of the triangle. A fair criteria to evaluate the quality of a mesh is to provide the minimum and mean values of the quality factor, cf. \cite{Alliez2005}. Here both are used as control parameters to determine when the iterative algorithm has reached the desired mesh quality tolerances (\cref{fig:flow_chart}).

\subsection*{Step 4: Local mesh improvements}
\label{Local mesh improvements}
So far the above algorithm would only move nodes within the domain to meet the desired spring lengths/internodal distances. However, in general we do not know a priori how many nodes are needed for a mesh. Therefore we use algorithms to locally add and remove nodes where the spacing is too loose or tight in the equilibrium configuration. After solving for nodal positions, we check if the mesh has reached the expected nodal density by determining the mean of the misfit in spring lengths (\cref{fig:flow_chart}). This is given by
\begin{equation}
  \label{eq:16}
  \mu
  =\frac{1}{N} \sum_{i=1}^N \left| \frac{l_i-{l_0}_i}{{l_0}_i} \right|
\end{equation}
where $l$ is the actual spring length, $l_0$ is the desired spring length and $N$ is the total number of springs in the mesh. Nodes are added or rejected (see below) if $\mu \geq \mu_t$. When $\mu < \mu_t$ the expected nodal density is achieved and element shape improvements (see below) are applied to obtain higher quality elements. After some experimentation we found it appropriate to use $0.02 < \mu_t < 0.05$ for 2-D meshes.

\textit{\textbf{Add/reject nodes}}. In the iterative process of mesh generation the possibility to either add or reject nodes plays an important local role. This feature is especially relevant when the goal is to create a global coarse mesh with an embedded high resolution sub-region. The logic for adding or rejecting nodes is based on the relative length change of the springs connecting nodes
\begin{equation}
  	\label{eq:17}
  	\epsilon
  	=\frac{l-l_0}{l_0}
\end{equation}
indicating whether springs are stretched $ \left( \epsilon > 0 \right) $ or compressed $ \left( \epsilon < 0 \right) $ with respect to their desired lengths. A new node is created at the midpoint of those springs with $ \epsilon > 0.5 $, i.e., springs stretched more than 50\% greater than their desired length. One node at the end of a spring is rejected when $ \epsilon < -0.5 $, i.e., springs compressed more than 50\% below their desired length. In order to save computational time, the add/reject nodes routine is called as a sub-iteration within the main iteration in which nodal positions are found. Sub-iterations are performed until the percentage of springs with $|\epsilon| > 0.5$ in the sub-iteration $j+1$ is higher than in the sub-iteration $j$. This implementation is especially useful when a large fraction of nodes need to be either added or rejected within a particular region of the mesh, e.g., when a relatively poor initial guess is used.

\begin{figure} [b]
	\centering
    \includegraphics[width=1\textwidth]{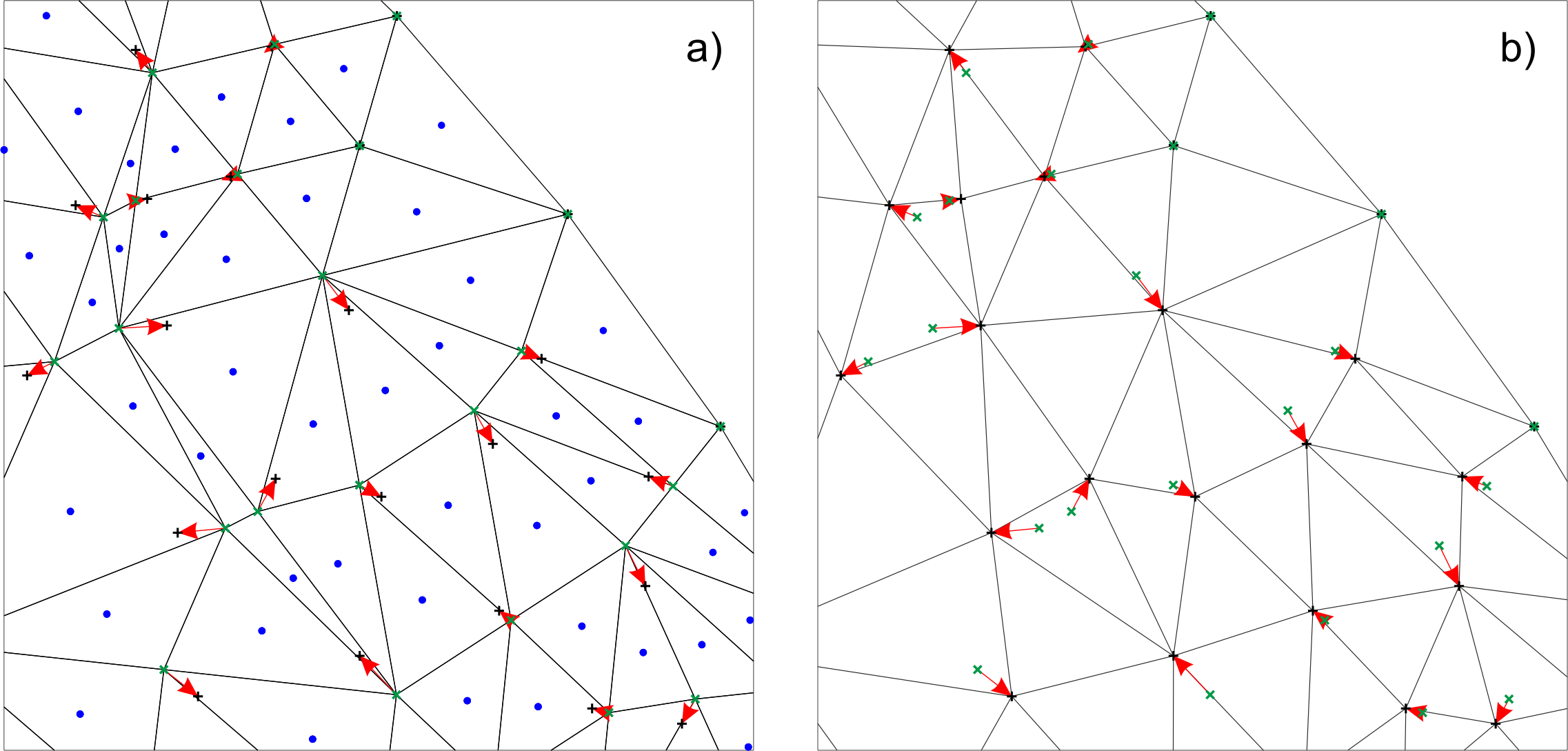}
    \caption{(a) Initial 2-D mesh. (b) Mesh after applying the Laplacian correction to smooth positions of its interior nodes. Blue points are the barycentres of the triangles. Green and black crosses are the nodal positions before and after smoothing, respectively. Red arrows indicate the motions of interior nodes.}
    \label{fig:smooth_int_nodes_2D_analogy_v2}
\end{figure}

\textit{\textbf{Smooth positions of the interior nodes}}. Good quality meshes are directly related to the generation of isotropic elements \cite{Alliez2005}. A Laplacian smoothing criteria, cf. \cite{Choi2003}, is used to improve the shape of poorly shaped elements, i.e., to make elements as close to a equilateral triangles or regular tetrahedra as possible. This method is only applied to interior nodes. The routine repositions interior nodes towards the mean of the barycentres of their surrounding elements, i.e.,
\begin{equation}
  \label{eq:18}
  \boldsymbol{x_s}
  =\frac{\sum\limits_{i=1}^N \boldsymbol{x_b}_i}{N} 
\end{equation}
where $\boldsymbol{x_s}$ are the new coordinates of the interior node, $N$ is the number of elements surrounding the interior node and $\boldsymbol{x_b}_i$ are the barycentre coordinates of the $i$-th surrounding element. \cref{fig:smooth_int_nodes_2D_analogy_v2} shows an example of smoothing positions of interior nodes for a 2-D mesh.

\subsection*{Example: Rectangular mesh with an embedded high resolution region}
\label{sec:Example: Rectangular mesh with an embedded high resolution region}
\begin{figure}
  	\centering
    \includegraphics[width=0.88\textwidth]{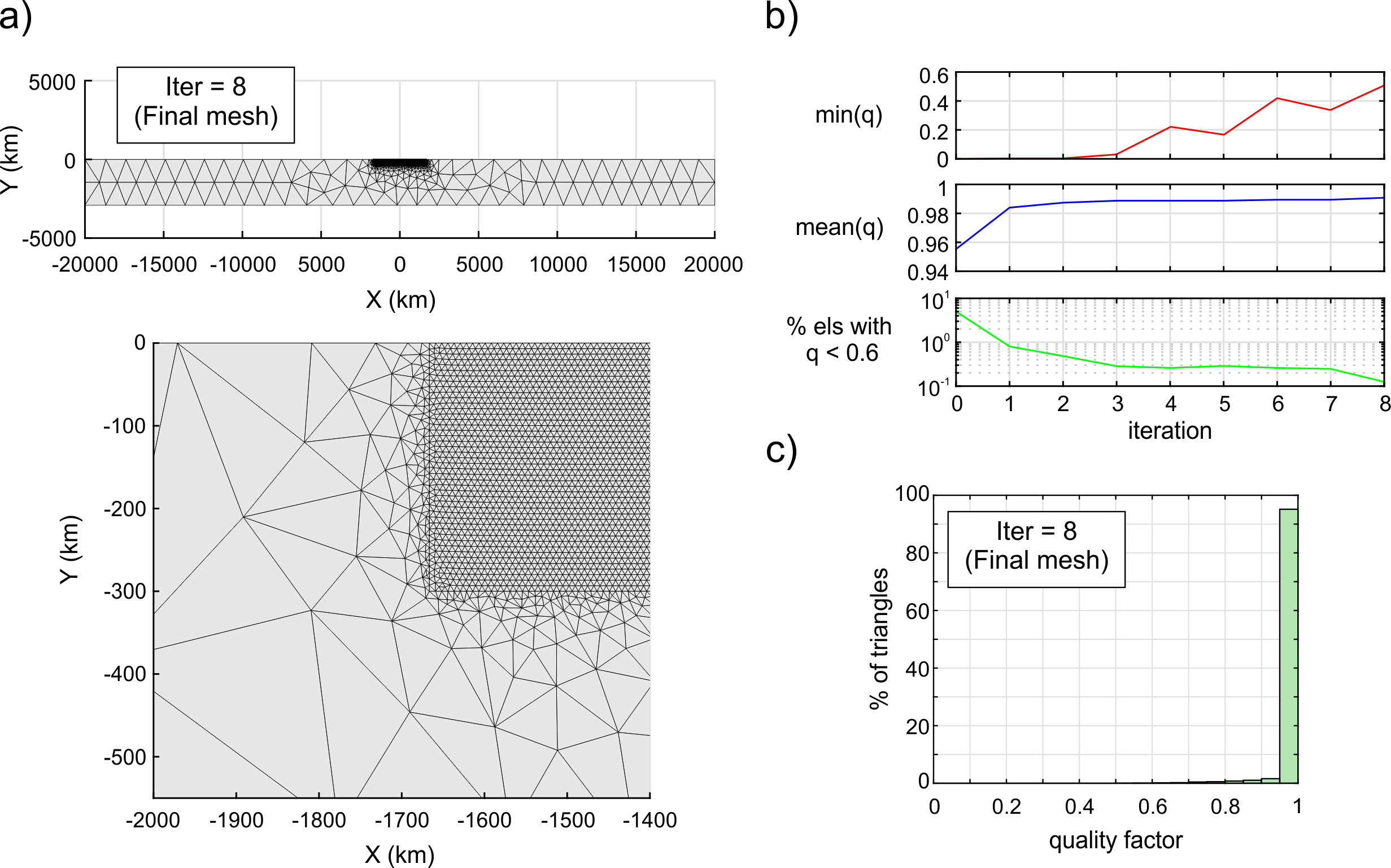}
  	\caption{(a) Final mesh (top) for a rectangular box with an embedded high resolution sub-region and a zoom around the left boundary of the refined region (bottom). (b) Minimum quality factor (red line), mean quality factor for all elements (blue line) and percentage of elements having a quality factor lower than 0.6\% (green line) as a function of iteration number. (c) Histogram of the fraction of elements as a function of quality factor for the final mesh.}
  	\label{fig:results_RECT_v4}
\end{figure}

Several tests have been performed with the above implementations in order to demonstrate the robustness of this mesh-generation recipe. As an example, we show the results for a rectangular box with an embedded high-resolution sub-region (code available in \cref{SM3}). The input parameters that control the algorithm are listed in \cref{tab:1}. The algorithm created the mesh in 9 s (all tests in this study have been performed using MATLAB R2015a (8.5.0.197613) on a 3.2 GHz Intel Core i5 (MacOSX 10.12.5) with 24 GB of 1600 MHz DDR3 memory) after eight outermost loop iterations (cf. \cref{fig:flow_chart}). \cref{fig:results_RECT_v4}a shows the final mesh (top) and a zoom around the left boundary of the refined region (bottom) for the iteration 8 (see \cref{fig:results_RECT_v4_SM} for iterations 0 (initial mesh) and 1). The final mesh has 22000 nodes forming 43000 triangles (\cref{tab:2}) with an edge-length factor $ l_{0r}/l_{0c} = 1/200$. The percentage of triangles within the coarse, transition and refined regions is 0.3\%, 6.3\% and 93.4\% respectively. The lowest quality factor for an element is 0.51 (red line in \cref{fig:results_RECT_v4}b) and the mean quality factor for all elements is 0.99 (blue line in \cref{fig:results_RECT_v4}b). Only 0.12\% of the triangles have a quality factor lower than 0.6 (green line in \cref{fig:results_RECT_v4}b). \cref{fig:results_RECT_v4}c shows the fraction of elements as a function of quality factor for the final mesh.
\begin{table}
	\caption{Mesh Parameters.}
	\label{tab:1}
	\centering
	{\small
	\begin{tabular}{ l p{5.1cm} r r r }
	\hline\noalign{\smallskip}
	Symbol & 
	Meaning & 
	\begin{tabular}{@{}c@{}}Rectangular \\ box\end{tabular} & 
	\begin{tabular}{@{}c@{}}Cylindrical \\ annulus\end{tabular} & 
	\begin{tabular}{@{}c@{}}Spherical \\ shell\end{tabular}     \\
	\noalign{\smallskip}\hline\noalign{\smallskip}\noalign{\smallskip}\noalign{\smallskip}
	$ d $		    & Depth														    & 2900 km       	& -             		& - 					\\
	$ l $ 			& Length															& 40000 km      	& -             		& - 					\\
	$ r_{i} $   		& Inner radius 													& -             	& 3471 km       		& 3471 km 	  		\\
	$ r_{o} $	   	& Outer radius													& -             	& 6371 km 	    		& 6371 km 	  		\\
	$ x_{0} $	 	& x-coordinate centre of refined region							& 0 km			& -             		& - 	        			\\
	$ z_{0} $	 	& z-coordinate centre of refined region		   					& 0 km			& -					& - 		    			\\
	$ \theta_0 $ 	& Colatitude centre of refined region	 	 	   				& -             	& $ 90^{\circ} $ 	& $ 90^{\circ} $ 	\\
	$ \phi_0 $ 		& Longitude centre of refined region		  	   					& -             	& -		 	    		& $ 90^{\circ} $ 	\\
	$ r_0 $ 			& Radial distance centre of refined region 	   					& -             	& 6371 km	    		& 6371 km   			\\
	$ l_{0 \ c} $ 	& Desired spring length for elements inside the coarse region  	& 1500 km       	& 2000 km       		& 2000 km   			\\
	$ l_{0 \ r} $ 	& Desired spring length for elements inside the refined region 	& 7.5 km        	& 10 km         		& 60 km 	  			\\
	$ d_{t} $ 	    & Transition region depth 										& 2900 km       	& 2900 km       		& 2900 km 	  		\\
	$ l_{t} $ 		& Transition region length 										& 8000 km       	& 8000 km       		& 6800 km   			\\
	$ w_{t} $ 		& Transition region width 										& -             	& -		 	    		& 9600 km   			\\
	$ d_{r} $ 	    & Refined region depth 											& 300 km 	  	& 300 km 	  		& 300 km 	  		\\
	$ l_{r} $ 	    & Refined region length 											& 3333 km   		& 3333 km   			& 2200 km   			\\
	$ w_{r} $ 	    & Refined region width 											& -             	& -		 	    		& 5000 km   			\\
	$ q_{t} $    	& Tolerance for minimum quality factor 							& 0.45 	  		& 0.30 	  			& 0.23 	  			\\
	$ \bar{q}_{t} $	& Tolerance for mean quality factor 								& 0.89 	  		& 0.93 	  			& 0.86 	  			\\
	$ {\mu}_{t} $ 	& Tolerance for mean misfit spring length 						& 0.025 	  		& 0.04 	  			& 0.11 	  			\\
	\noalign{\smallskip}\hline
	\end{tabular}
	}
\end{table}
\begin{table}
	\caption{Information on example meshes.}
	\label{tab:2}
	\centering
	{\small
	\begin{tabular}{ l r r r r r r}
	\hline\noalign{\smallskip}
	Mesh 													& 
	nodes 													& 
	elements 												& 
	time (s) 												& 
	iterations 												& 
	\begin{tabular}{@{}c@{}}time per \\ node\end{tabular}    & 
	\begin{tabular}{@{}c@{}}time per \\ element\end{tabular} \\
	\noalign{\smallskip}\hline
	\begin{tabular}{@{}l@{}}Rectangular \\ box\end{tabular}     	& 22000	&  43000 &   9 &  8 & $4.1\cdot10^{-4}$ & $2.1\cdot10^{-4}$ \\
	\begin{tabular}{@{}l@{}}Cylindrical \\ annulus\end{tabular} 	& 12000	&  23000 &  17 &  5 & $1.4\cdot10^{-3}$ & $7.4\cdot10^{-4}$ \\
	\begin{tabular}{@{}l@{}}Spherical \\ shell\end{tabular}		& 27000 	& 150000 & 224 & 10 & $8.3\cdot10^{-3}$ & $1.5\cdot10^{-3}$ \\
	\noalign{\smallskip}\hline
	\end{tabular}
	}
\end{table}

\section{2-D Cylindrical annulus work flow}
\label{sec:2-D Cylindrical annulus work flow}
The algorithm presented above needs to be slightly modified to generate a cylindrical annulus mesh. The white and orange boxes in \cref{fig:flow_chart} show the flowchart that describes this modified algorithm. Since the general algorithm is the same, in this section we only discuss the parts that differ from the rectangular mesh generator described previously. 

\subsection*{Cylindrical annulus guide-mesh}
\label{sec:Cylindrical annulus guide-mesh}
\begin{figure}
    \centering
    \includegraphics[width=1\textwidth]{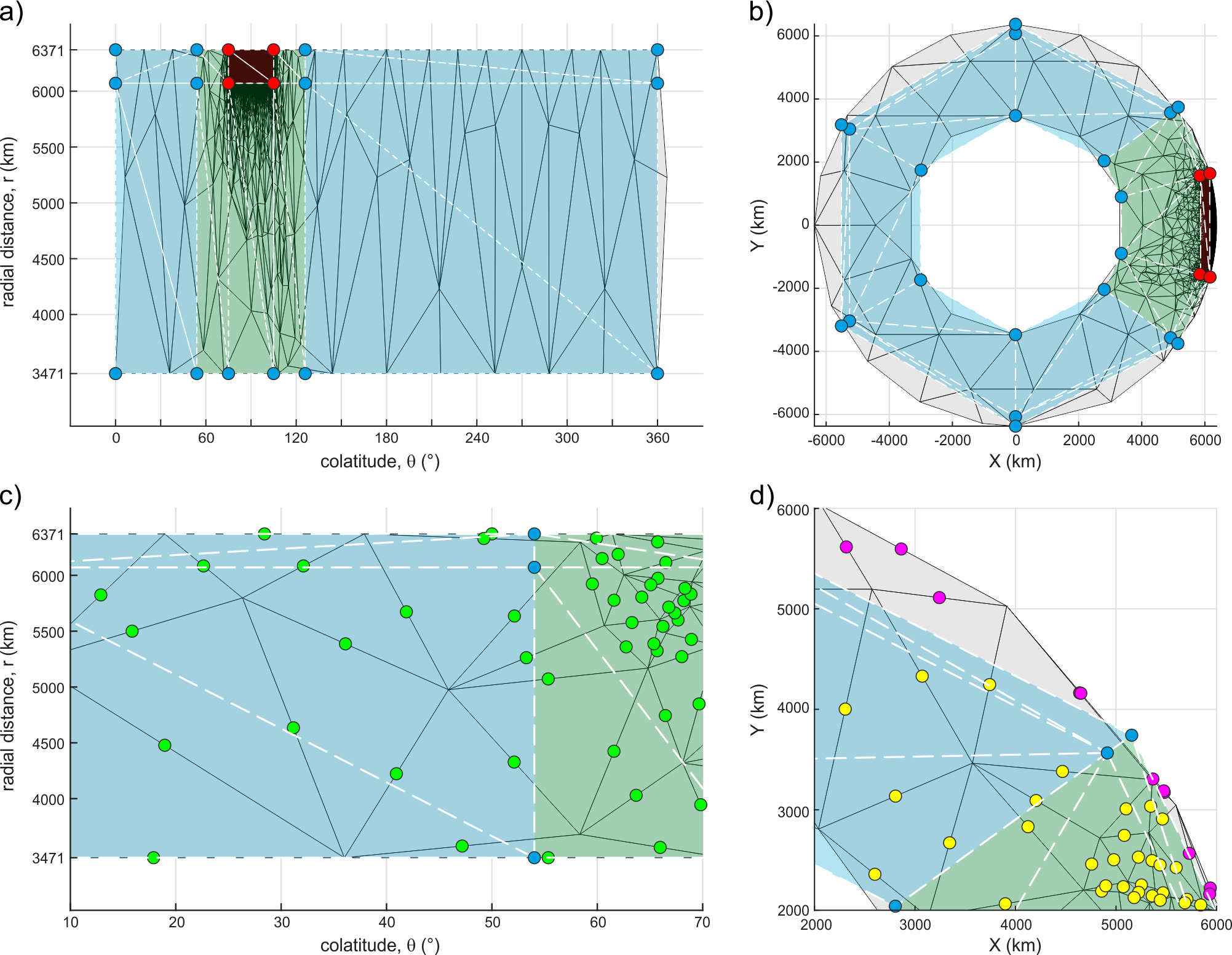}
    \caption{(a) Guide-mesh (white dashed lines) defined by a few nodes (red and blue dots represent $l_{0 \ r}$ and $l_{0 \ c}$ respectively) in polar coordinates for a cylindrical annulus mesh (initial guess is shown in black solid lines). Red, green and blue colours represent the refined, transition and coarse regions respectively. (b) Guide-mesh defined in Cartesian coordinates. Same colours as in (a). (c) Zoom around an edge of the transition region in polar coordinates. The function $l_0(\theta$, $r)$ can be interpolated at green dots with maximum precision since both boundaries -- the cylindrical annulus mesh and its guide-mesh -- are overlapping. (d) Zoom around an edge of the transition region in Cartesian coordinates. The function $l_0(x$, $y)$ cannot be interpolated at magenta dots since they lay outside of the outer boundary of a Cartesian guide-mesh. The precision of the interpolated $l_0$ values at yellow dots is reduced since both boundaries -- the cylindrical annulus mesh and its guide-mesh -- do not overlap.}
    \label{fig:guide_mesh_CYL_CART_vs_POL}
\end{figure}

The generation of a refined cylindrical annulus mesh using the guide-mesh involves the same steps as for a rectangular mesh except that the function $l_0(x$, $y)$ becomes $l_0(\theta$, $r)$. In this case the guide-mesh is a coarse cylindrical annulus mesh defined in polar coordinates. \cref{fig:guide_mesh_CYL_CART_vs_POL}a shows the guide-mesh (white dashed lines) defining the refined (red), transition (green) and coarse (blue) regions and the parameters are listed in \cref{tab:1}. Red and blue dots represent $l_{0 \ r}$ and $l_{0 \ c}$ respectively. The initial triangulation is shown in black solid lines. \cref{fig:guide_mesh_CYL_CART_vs_POL}c shows a zoom of the guide-mesh defined in polar coordinates. Green dots represent the points where the function $l_0(\theta$, $r)$ is interpolated. The use of a guide-mesh defined in polar coordinates (white dashed lines in \cref{fig:guide_mesh_CYL_CART_vs_POL}a and \cref{fig:guide_mesh_CYL_CART_vs_POL}c) instead of Cartesian coordinates (white dashed lines in \cref{fig:guide_mesh_CYL_CART_vs_POL}b and \cref{fig:guide_mesh_CYL_CART_vs_POL}d) takes advantage of higher precision when $l_0$ values are interpolated in points both close and on the boundaries (green dots in \cref{fig:guide_mesh_CYL_CART_vs_POL}c). This is because the shapes of the outer and inner boundaries of any cylindrical annulus mesh defined in Cartesian coordinates is not perfectly circular (\cref{fig:guide_mesh_CYL_CART_vs_POL}b). Therefore, it may occur that some boundary points (magenta dots in \cref{fig:guide_mesh_CYL_CART_vs_POL}d) may lay outside of the boundaries of a Cartesian guide-mesh (which can be a very coarse mesh) preventing accurate interpolation for the desired length at those points. Furthermore, the fact that both boundaries -- the cylindrical annulus mesh and its guide-mesh -- would not overlap in a Cartesian geometry would reduce the precision of the interpolated $l_0$ values (yellow dots in \cref{fig:guide_mesh_CYL_CART_vs_POL}d).

\subsection*{Circular Boundary Conditions}
\label{sec:Circular Boundary Conditions}
\begin{figure}
  	\centering
  	\includegraphics[width=0.8\textwidth]{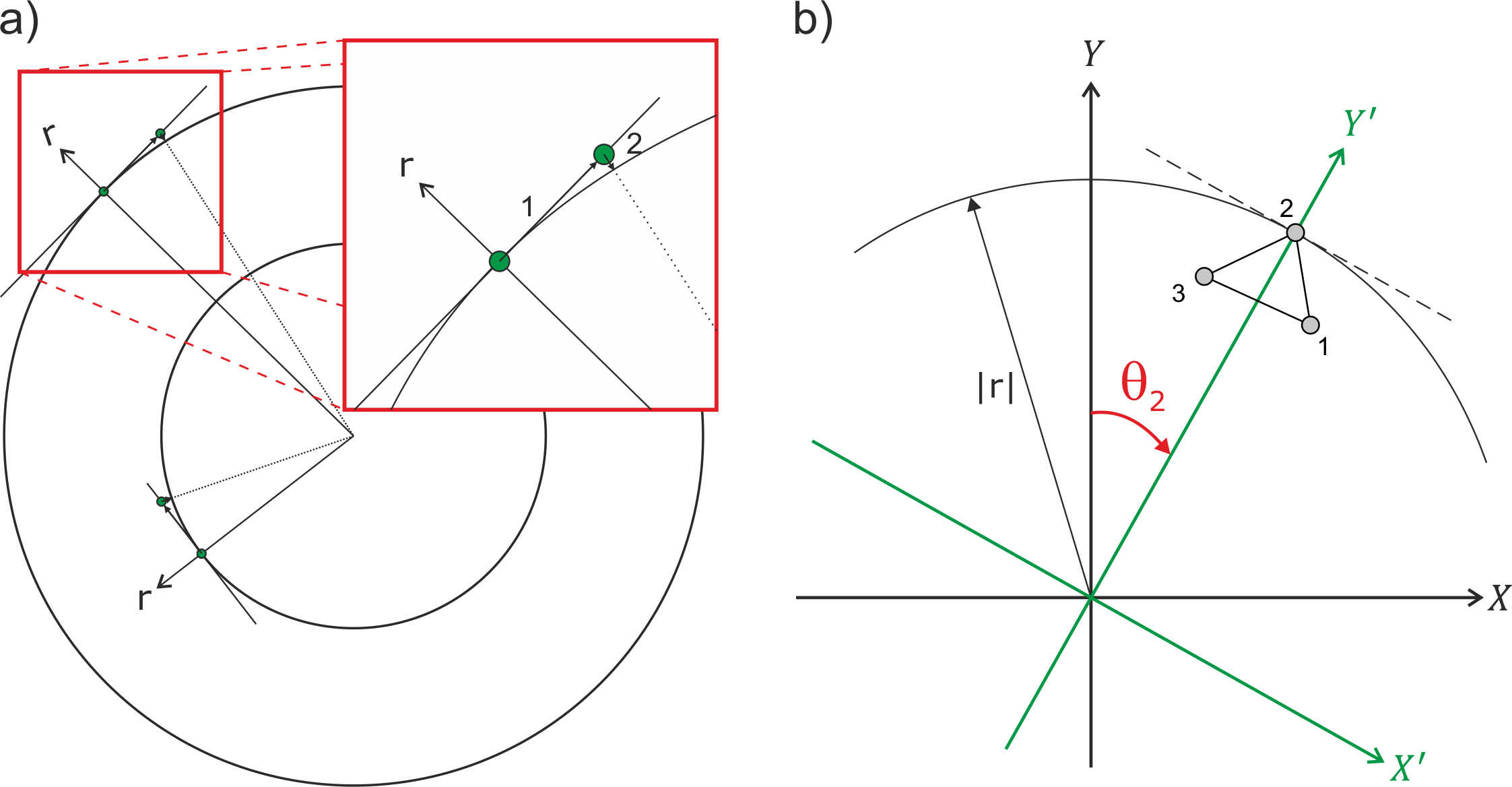}
  	\caption{(a) Conceptual diagram for circular boundary conditions. The motion of boundary nodes is first restricted to be along the tangent line to the circle. Then they are 'pulled back' to the circle by projecting in the radial direction. (b) Implementation of circular boundary conditions for one triangle. A rotation is needed for the node 2 in order to pass from the global reference system ($X$, $Y$) to the local surface-parallel reference system ($X'$, $Y'$) where $ {y_2}'=\vert r\vert $ is the constrained boundary condition.}
    \label{fig:BCs_CYL}
\end{figure}

Boundary conditions for a cylindrical annulus mesh are a generalization to the treatment for a straight-sided boundary line-segment. We denote the inner and outer boundaries $ \Sigma $ of the cylindrical annulus mesh as radii $r = r_{inner}$ and $r = r_{outer} $ respectively. $ \Omega $ is the interior region confined between both boundaries. A useful boundary condition is to prescribe nodes on $\Sigma$ that are free to move along the circular boundary. This nodal motion is generated by two independent steps (\cref{fig:BCs_CYL}a): 1) The node is allowed to move along the tangent line to the circle at its current location, and 2) the node is place onto the circle by projecting its new location in the radial direction. This approximation assumes that the radial distance needed to put the node back onto the circle is small compared to the distance moved along the tangent line.
For simplicity, the mathematical implementation is presented here only for one triangle (\cref{fig:BCs_CYL}b). The boundary condition for node 2 is that it slides along its tangent line (dashed line in Figure~\ref{fig:BCs_CYL}b) since $ {y_2}'=\vert r\vert $, where $ r $ is the radial distance from the centre of the cylindrical annulus mesh to the boundary. The boundary condition is imposed by a rotation of the coordinate system for node 2 given by the transformation matrix $ \boldsymbol{T} $ that relates global coordinates $ \boldsymbol{x} $ with local coordinates $ \boldsymbol{x}' $ (local surface-parallel reference system ($X'$, $Y'$) in green in \cref{fig:BCs_CYL}b) by
\begin{equation}
\label{eq:19}
	\underbrace{\left(\begin{array}{c}
		x_1 \\
		y_1 \\
		x_2 \\
		y_2 \\
		x_3 \\
		y_3
	\end{array}\right)}_{\boldsymbol{x}}
	=
	\underbrace{\left[\begin{array}{rrrrrr}
	1\  								       \\
	\ & 1\  							       \\
    \ & \ &  \cos{\theta_2} & \sin{\theta_2} \ \\
	\ & \ & -\sin{\theta_2} & \cos{\theta_2} \ \\
	\ & \ & \ & \ & 1\ \   			           \\
	\ & \ & \ & \ & \ & 1\ \
	\end{array}\right]}_{\boldsymbol{T}}
	\underbrace{\left(\begin{array}{c}
		x_1    		 \\
		y_1    		 \\
		{x_2}' 		 \\
        \vert r\vert \\
		x_3    		 \\
		y_3
	\end{array}\right)}_{\boldsymbol{x}'}
\end{equation}
where $ \theta_2 $ is the angle of the node 2 measured from the $ Y $ axis in the clockwise direction. After applying the transformation matrix to the stiffness matrix and force vector
\begin{equation}
  	\label{eq:20}
    \boldsymbol{K}' = \boldsymbol{T}^{\mkern-1.5mu\mathsf{T}} \boldsymbol{K} \boldsymbol{T}
\end{equation}
\begin{equation}
    \label{eq:21}
    \boldsymbol{f_{l_0}}' = \boldsymbol{T}^{\mkern-1.5mu\mathsf{T}} \boldsymbol{f_{l_0}}
\end{equation}
the new system of equations is given by
\begin{equation}
    \label{eq:22}
    \boldsymbol{K}' \boldsymbol{x}' = \boldsymbol{f_{l_0}}'
\end{equation}
which is then solved for $ \boldsymbol{x}' $ . Global coordinates are recovered through the transformation matrix
\begin{equation}
    \label{eq:23}
    \boldsymbol{x} = \boldsymbol{T} \boldsymbol{x}'
\end{equation}

\subsection*{Add/reject nodes in cylindrical annulus meshes}
\label{sec:Add-reject nodes in cylindrical annulus meshes}
The routine to add or reject nodes for a cylindrical annulus mesh works like the one explained above for a rectangular mesh. The only difference appears when a new node is added on a boundary spring. In this case, the new boundary node needs to be projected onto the surface along the radial direction.

\subsection*{Example: Cylindrical annulus mesh with an embedded high resolution region}
\label{sec:Example: Cylindrical annulus mesh with an embedded high resolution region}
\begin{figure} [t]
  	\centering
    \includegraphics[width=0.88\textwidth]{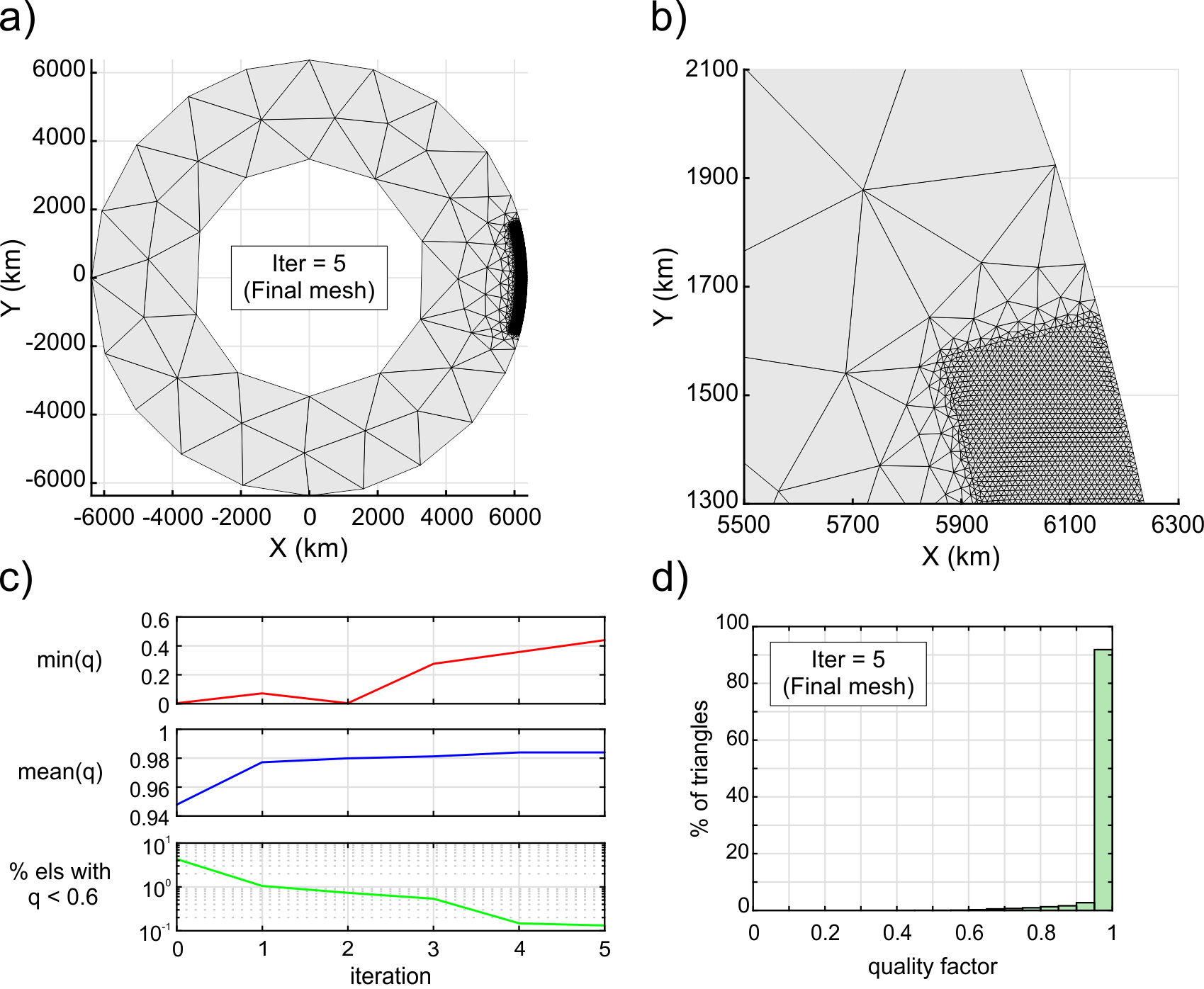}
  	\caption{(a) Final mesh for a cylindrical annulus with an embedded high resolution sub-region. (b) Zoom around an edge of the refined region. (c) Minimum quality factor (red line), mean quality factor for all elements (blue line) and percentage of elements having a quality factor lower than 0.6\% (green line) as a function of iteration number. (d) Histogram of the fraction of elements as a function of quality factor for the final mesh.}
  	\label{fig:results_CYL_v4}
\end{figure}
We show the results for a cylindrical annulus mesh with an embedded high-resolution sub-region (code available in \cref{SM4}). The input generation parameters are listed in \cref{tab:1}. The algorithm created the mesh in 17 s after 5 iterations. \cref{fig:results_CYL_v4}a shows the final mesh (top) and a zoom around an edge of the refined region (bottom) for iteration 5 (see \cref{fig:results_CYL_v4_SM} for iterations 0 (initial mesh) and 1). The final mesh has 12000 nodes forming 23000 triangular elements (\cref{tab:2}) with an edge-length factor $ l_{0r}/l_{0c} = 1/200$. The percentage of triangles within the coarse, transition and refined regions is 0.2\%, 6.0\% and 93.8\% respectively. The worst quality factor for an element is 0.44 (red line in \cref{fig:results_CYL_v4}b) and the mean quality factor of all elements is 0.98 (blue line in \cref{fig:results_CYL_v4}b). Only 0.13\% of the triangles have a quality factor lower than 0.6 (green line in \cref{fig:results_CYL_v4}b). \cref{fig:results_CYL_v4}c shows the fraction of elements as a function of their quality factor for the final mesh.

\section{3-D Spherical shell work flow}
\label{sec:3-D Spherical shell work flow}
The algorithm presented above was developed as an intermediate step towards the generation of 3-D spherical shell meshes that include an embedded high resolution sub-region. The white and green backgrounds in \cref{fig:flow_chart} show the flowchart that describes the 3-D spherical algorithm. In this section we discuss those parts of the algorithm that differ from the cylindrical annulus mesh generator.

\subsection*{Initial placement of the nodes in 3-D}
\label{sec:Initial placement of the nodes in 3-D}

The boundary nodes in the refined and coarse regions are created by recursively splitting an initial dodecahedron according to $l_{0r}$ and $l_{0c}$ respectively. This gives a uniform distribution of equilateral triangles on the spherical surface. In contrast to equilateral triangles in 2-D, which are able to fill up the plane, regular tetrahedra do not fill up the entire space. However, there do exist some compact lattices, e.g., the hexagonal close packing (hcp) lattice, that create a distribution of nodes that leads to well shaped tetrahedra. The interior nodes within the refined and coarse regions are created by a close-packing of equal spheres with radii equal to $l_{0r}/2$ and $l_{0c}/2$ respectively. The initial placement for boundary and interior nodes in the transition region is created using $l_{0r}$ as explained above. Then the rejection method described in \cite{Persson2004} is used to discard points and create a weighted distribution of nodes.

\subsection*{Spherical shell guide-mesh}
\label{sec:Spherical shell guide-mesh}
\begin{figure}
    \centering
    \includegraphics[width=1\textwidth]{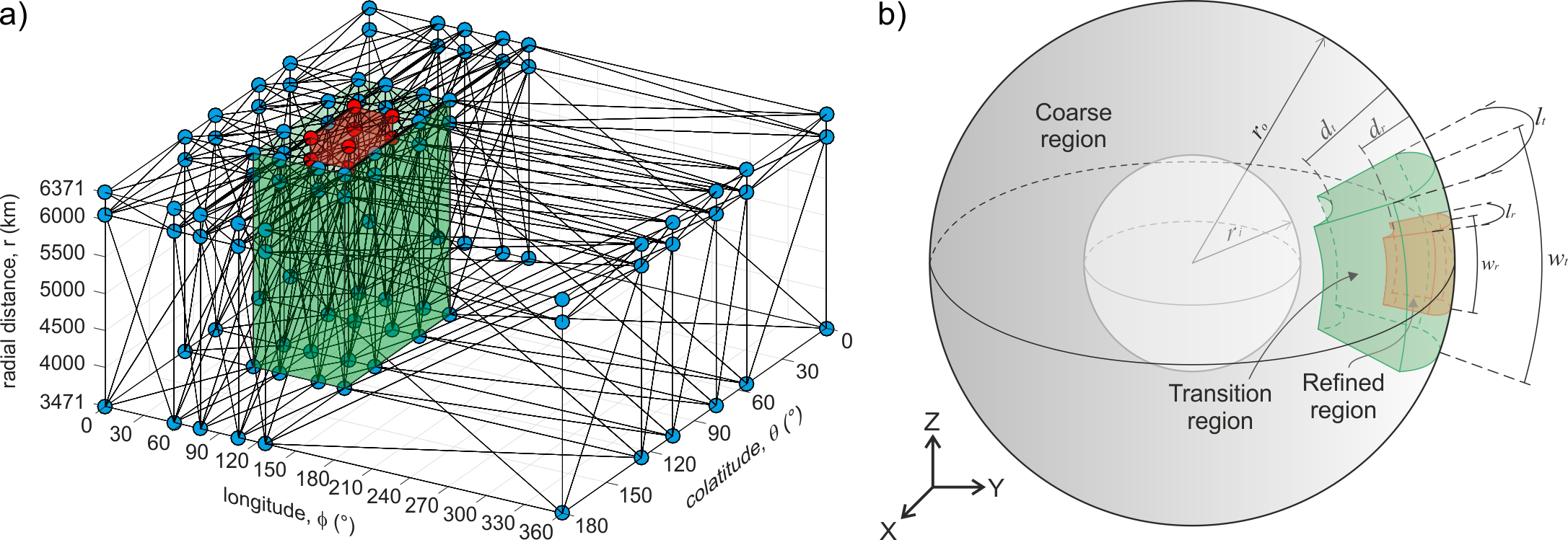}
    \caption{(a) Guide-mesh defined by a few nodes (red and blue dots represent $l_{0r}$ and $l_{0c}$ respectively) in spherical coordinates for a spherical shell. The length of the springs within the refined region (red) is approximately equal to $l_{0r}$. The length of the springs within the transition region (green) smoothly varies from $l_{0r}$ to $l_{0c}$. Outside the transition region the length of the springs is approximately equal to $l_{0c}$. (b) Model domain representing a 3-D spherical shell with an embedded high resolution sub-region.}
    \label{fig:guide_mesh_SPH_ans_mesh_setup}
\end{figure}

The generation of a refined spherical shell mesh using the guide-mesh involves steps similar to those described above except that the preferred length function $l_0(x$, $y)$ is now $l_0(\theta$, $\phi$, $r)$. In this case the guide-mesh is a coarse spherical shell mesh defined in spherical coordinates (Figure~\ref{fig:guide_mesh_SPH_ans_mesh_setup}a).

\subsection*{Spring-based solver in 3-D}
\label{sec:Spring-based solver in 3-D}
The spring-based solver described above naturally extends to 3-D. Forces and nodal positions are expressed in $x$, $y$ and $z$ coordinates (\cref{fig:springs_2D_and_3D}b). In order to solve for nodal positions in 3-D, a change from local coordinates (${x_1}'$, $0$, $0$; ${x_2}'$, $0$, $0$) to global coordinates ($x_1$, $y_1$, $z_1$; $x_2$, $y_2$, $z_2$) is needed. This change of coordinates consists of a 3-D rotation described by the rotation matrix
\begin{equation}
  	\label{eq:24}
  	\boldsymbol{R_{3D}} =
  	\left[\begin{array}{cccccc}
  	\cos\alpha\cos\beta & \cos\alpha\sin\beta & \sin\alpha &                   0 &                   0 &          0 \\
  	                  0 &                   0 &          0 & \cos\alpha\cos\beta & \cos\alpha\sin\beta & \sin\alpha
  	\end{array}\right]
\end{equation}
where $\alpha$ and $\beta$ are angles equivalents to latitude and longitude, respectively (\cref{fig:springs_2D_and_3D}b). Applying equation \cref{eq:24} to equation \cref{eq:3} (see \cref{SM2} for details), equation \cref{eq:3} becomes
\begin{equation}
	\label{eq:25}
	{\footnotesize
	\begin{split}
	& k	
	\left[\begin{array}{cccccc}
		{-c_\alpha}^2 {c_\beta}^2     & {-c_\alpha}^2 s_\beta c_\beta & -s_\alpha c_\alpha c_\beta 	&
		 {c_\alpha}^2 {c_\beta}^2     &  {c_\alpha}^2 s_\beta c_\beta &  s_\alpha c_\alpha c_\beta 	\\
		{-c_\alpha}^2 s_\beta c_\beta & {-c_\alpha}^2 {s_\beta}^2     & -s_\alpha c_\alpha s_\beta 	&
		 {c_\alpha}^2 s_\beta c_\beta &  {c_\alpha}^2 {s_\beta}^2     &  s_\alpha c_\alpha s_\beta 	\\
		-s_\alpha c_\alpha c_\beta    & -s_\alpha c_\alpha s_\beta    & {-s_\alpha}^2 			   	&
		 s_\alpha c_\alpha c_\beta    &  s_\alpha c_\alpha s_\beta    &  {s_\alpha}^2 		       	\\
		 {c_\alpha}^2 {c_\beta}^2     &  {c_\alpha}^2 s_\beta c_\beta &  s_\alpha c_\alpha c_\beta 	&
		{-c_\alpha}^2 {c_\beta}^2     & {-c_\alpha}^2 s_\beta c_\beta & -s_\alpha c_\alpha c_\beta 	\\
		 {c_\alpha}^2 s_\beta c_\beta &  {c_\alpha}^2 {s_\beta}^2     &  s_\alpha c_\alpha s_\beta 	&
		{-c_\alpha}^2 s_\beta c_\beta & {-c_\alpha}^2 {s_\beta}^2     & -s_\alpha c_\alpha s_\beta 	\\
		 s_\alpha c_\alpha c_\beta    &  s_\alpha c_\alpha s_\beta    &  {s_\alpha}^2 			   	&
		-s_\alpha c_\alpha c_\beta    & -s_\alpha c_\alpha s_\beta    & {-s_\alpha}^2 			 
	\end{array}\right] 
	\left(\begin{array}{c}
		x_1 \\
		y_1 \\
		z_1 \\
		x_2 \\
		y_2 \\
		z_2 \\
	\end{array}\right) \\
    &
	=
	\left(\begin{array}{c}
		f_{1,x} \\
		f_{1,y} \\
		f_{1,z} \\
		f_{2,x} \\
		f_{2,y} \\
		f_{2,z} \\
	\end{array}\right)
	+
	kl_0
	\left(\begin{array}{c}
		 c_\alpha c_\beta \\
		 c_\alpha s_\beta \\
		 s_\alpha         \\
		-c_\alpha c_\beta \\
		-c_\alpha s_\beta \\
		-s_\alpha
	\end{array}\right)
	\end{split}
	}
\end{equation}
where $s_\alpha\equiv\sin\alpha$, $c_\alpha\equiv\cos\alpha$, $s_\beta\equiv\sin\beta$ and $c_\beta\equiv\cos\beta$. The system of equations is solved as described above (see equation \cref{eq:7}).

\subsection*{Spherical Boundary Conditions}
\label{sec:Spherical Boundary Conditions}
\begin{figure} [ht]
  	\centering
  	\includegraphics[width=0.8\textwidth]{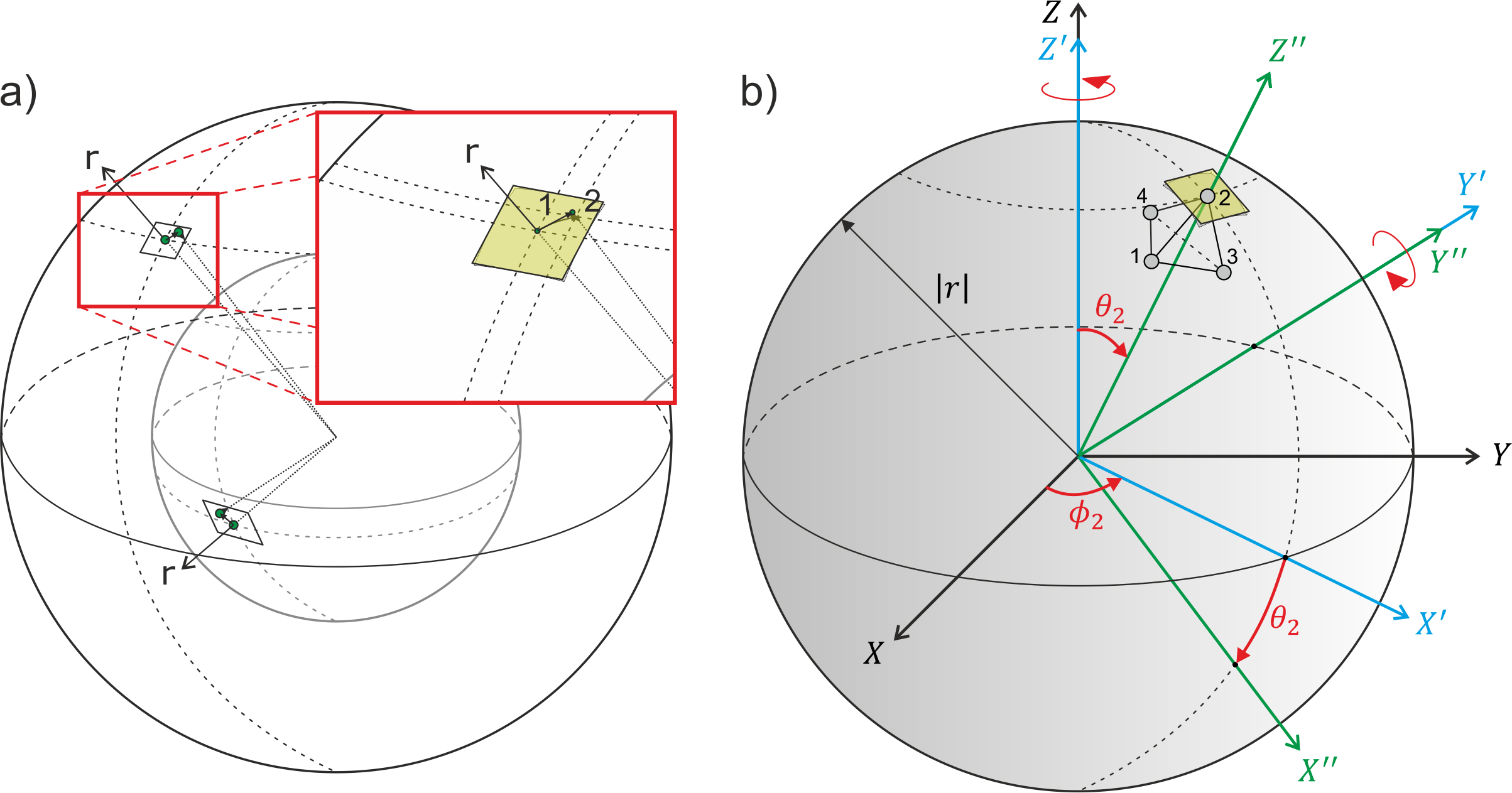}
  	\caption{(a) Conceptual diagram for spherical boundary conditions. The motion of boundary nodes is first restricted to be along the tangent plane to the sphere. Then, they are 'pulled back' to the sphere's surface by projecting in the radial direction. (b) Implementation of spherical boundary conditions for one tetrahedron. Two rotations are needed for node 2 to pass from the global reference system ($X$, $Y$, $Z$) to the local reference system ($X''$, $Y''$, $Z''$), where $ {z_2}''=\vert r\vert $ is the boundary condition.}
    \label{fig:BCs_SPH}
\end{figure}

For 3-D applications, we currently focus on developing unstructured spherical meshes. Using a notation similar to that for 2-D circular boundary conditions, we denote by $ \Sigma $ the inner and outer boundaries of the spherical shell with radii $r = r_{inner}$ and $r = r_{outer} $ respectively. $ \Omega $ is the interior region between the boundaries. A useful boundary condition consists in prescribing boundary nodes that are free to slide along the local tangent plane to the spherical surface. Nodal sliding is generated in two independent steps (\cref{fig:BCs_SPH}a): 1) The node is allowed to move along the local tangent plane to the sphere, and 2) the node is returned to the sphere's surface by projecting in the radial direction. This approximation assumes that the radial distance needed to pull the node back to the surface of the sphere is small compared to the distance moved along the tangent plane.
For simplicity, the mathematical implementation of the spherical boundary conditions is presented here only for one tetrahedron (\cref{fig:BCs_SPH}b). Node 2 is free to slide along the tangent plane since the boundary condition is $ {z_2}''=\vert r\vert $, where $ r $ is the radial distance from the centre of the sphere to the surface. This boundary condition is imposed by two rotations of the coordinate system for node 2. The first rotation is around the $ Z $ axis by an angle $ \phi_2 $, which is the longitude of node 2 (local reference system $(X',Y',Z')$ in blue in \cref{fig:BCs_SPH}b). The second rotation is around the $ Y' $ axis by an angle $ \theta_2 $, which is the colatitude for node 2 (local reference system $(X'',Y'',Z'')$ in green in \cref{fig:BCs_SPH}b). The complete rotation is given by the transformation matrix $ \boldsymbol{T} $ that relates global coordinates $ \boldsymbol{x} $ with local coordinates $ \boldsymbol{x}'' $ as follows
\begin{equation}
\label{eq:26}
	{\scriptsize
	\underbrace{\left(\begin{array}{c}
		x_1 \\
		y_1 \\
		z_1 \\
		x_2 \\
		y_2 \\
		z_2 \\
		x_3 \\
		y_3 \\
		z_3 \\
		x_4 \\
		y_4 \\
		z_4
	\end{array}\right)}_{\boldsymbol{x}}
	=
	\underbrace{\left[
	\arraycolsep=1.4pt
	\begin{array}{rrrrrrrrrrrr}
	1\ \ \\
	\ & 1\ \ \\
	\ & \ & 1\ \ \\
	\ & \ & \ &  \cos{\phi_2}\cos{\theta_2}\ & -\sin{\phi_2}\   & \cos{\phi_2}\sin{\theta_2} \\
	\ & \ & \ &  \sin{\phi_2}\cos{\theta_2}\ &  \cos{\theta_2}\ & \sin{\phi_2}\sin{\theta_2} \\
	\ & \ & \ & -\sin{\theta_2}\             &  0\              & \cos{\theta_2}			 \\
	\ & \ & \ & \ & \ & \ & 1\ \ \\
	\ & \ & \ & \ & \ & \ & \ & 1\ \ \\
	\ & \ & \ & \ & \ & \ & \ & \ & 1\ \ \\
	\ & \ & \ & \ & \ & \ & \ & \ & \ & 1\ \ \\
	\ & \ & \ & \ & \ & \ & \ & \ & \ & \ & 1\ \ \\
	\ & \ & \ & \ & \ & \ & \ & \ & \ & \ & \ & 1\ \
	\end{array}
	\right]}_{\boldsymbol{T}}
	\underbrace{\left(\begin{array}{c}
		x_1 \\
		y_1 \\
		z_1 \\
		{x_2}'' \\
		{y_2}'' \\
		\vert r\vert \\
		x_3 \\
		y_3 \\
		z_3 \\
		x_4 \\
		y_4 \\
		z_4
	\end{array}\right)}_{\boldsymbol{x}''}
	}
\end{equation}
This transformation matrix contains a $ \theta $ and $ \phi $ angle for each node on the spherical boundary. Applying the transformation matrix to stiffness matrix and force vector
\begin{equation}
  	\label{eq:27}
    \boldsymbol{K}'' = \boldsymbol{T}^{\mkern-1.5mu\mathsf{T}} \boldsymbol{K} \boldsymbol{T}
\end{equation}
\begin{equation}
    \label{eq:28}
    \boldsymbol{f_{l_0}}'' = \boldsymbol{T}^{\mkern-1.5mu\mathsf{T}} \boldsymbol{f_{l_0}}
\end{equation}
the new system of equations is given by
\begin{equation}
    \label{eq:29}
    \boldsymbol{K}'' \boldsymbol{x}'' = \boldsymbol{f_{l_0}}''
\end{equation}
which is solved for $ \boldsymbol{x}'' $. Global Cartesian coordinates are recovered through the transformation matrix
\begin{equation}
    \label{eq:30}
    \boldsymbol{x} = \boldsymbol{T} \boldsymbol{x}''
\end{equation}

\subsection*{Quality factor for tetrahedra}
\label{sec:Quality factor for tetrahedra}
\begin{figure}
    \centering
    \includegraphics[width=0.6\textwidth]{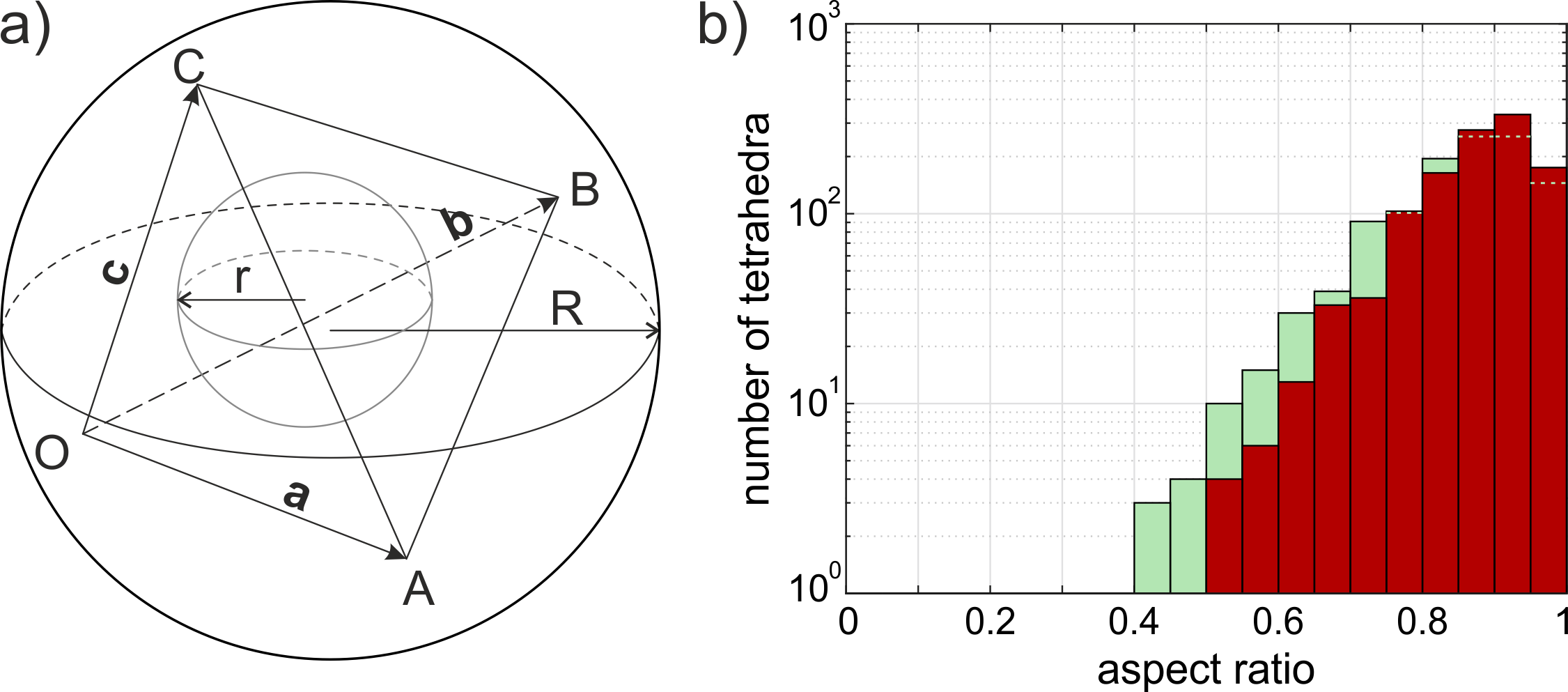}
    \caption{(a) Tetrahedron with vertices OABC. $ R $ and $ r $ are the radius of the circumscribed and inscribed spheres respectively. (b) Number of tetrahedra as a function of the quality factor $ q_{_{3D}} $ (green) and the shape measure $ s $ (red) for the same mesh.}
    \label{fig:q_factor_and_q_vs_s}
\end{figure}

The 3-D quality factor for a tetrahedron is defined by
\begin{equation}
    \label{eq:31}
    q_{_{3D}} = \frac{3r_s}{R_s}
\end{equation}
where $ r_s $ is the radius of the tetrahedron's inscribed sphere and $ R_s $ is the radius of its circumscribed sphere. $ R_s $ and $ r_s $ are given by
\begin{equation}
    \label{eq:32}
    r_s
    =
    \frac{| \boldsymbol{a} \cdot \left( \boldsymbol{b} \times \boldsymbol{c} \right) |}
         {\left( | \boldsymbol{a} \times \boldsymbol{b} | +
                 | \boldsymbol{b} \times \boldsymbol{c} | + 
                 | \boldsymbol{c} \times \boldsymbol{a} | +
                 | \left( \boldsymbol{a} \times \boldsymbol{b} \right) +
                   \left( \boldsymbol{b} \times \boldsymbol{c} \right) +
                   \left( \boldsymbol{c} \times \boldsymbol{a} \right) | 
          \right)}
\end{equation}
\begin{equation}
    \label{eq:33}
    R_s
    =
    \frac{| \boldsymbol{a}^2 \cdot \left( \boldsymbol{b} \times \boldsymbol{c} \right) + 
            \boldsymbol{b}^2 \cdot \left( \boldsymbol{c} \times \boldsymbol{a} \right) +
            \boldsymbol{c}^2 \cdot \left( \boldsymbol{a} \times \boldsymbol{b} \right)
          |}
         {2| \boldsymbol{a} \cdot \left( \boldsymbol{b} \times \boldsymbol{c} \right) |}
\end{equation}
where $ \boldsymbol{a} $, $ \boldsymbol{b} $ and $ \boldsymbol{c} $ are vectors pointing from one node, O, to the three other nodes of the tetrahedron $ A $, $ B $ and $ C $ respectively (Figure~\ref{fig:q_factor_and_q_vs_s}a). This quality factor is normalized to be 0 for degenerate tetrahedra and 1 for regular tetrahedra. Note that different definitions for normalized aspect ratios can lead to different estimators for the global quality of a mesh. For example, \cite{Anderson2005} define a shape measure $ s $ that depends on tetrahedral volume and the lengths of its edges. Computing $ q_{_{3D}} $ and $ s $ for the same mesh gives differences of up to 0.1 for the worst element (\cref{fig:q_factor_and_q_vs_s}b). The quality factor $ q_{_{3D}} $ that we choose to use is a more restrictive aspect ratio than the shape factor measure $ s $. 

\subsection*{Element shape improvements}
\label{sec:Element shape improvements}
In 3-D, even when the expected nodal density is achieved ($\mu < \mu_{t}$) by adding or rejecting nodes, a considerable number of poorly shaped tetrahedra can still persist. Local improvements are needed to ensure that the mesh is robust enough to perform optimal FEM calculations. After some experimentation, we found it appropriate to use $\mu_{t} = 0.11$ although this can vary from 0.1 to 0.2 depending on the degree of mesh refinement. The value of $\mu_{t}$ for 2-D meshes is smaller than for 3-D meshes due to the shape compactness that can be achieved on a 2-D planar surface.

Methods based on swapping edges or faces to improve element quality can possibly generate non-Delaunay triangulations, which will cause problems in algorithms that rely on a mesh created by a Delaunay triangulation (e.g. point search algorithms). Hence, as an alternative and in addition to smoothing the position of interior nodes, we recommend two additional operations to improve the quality of tetrahedral elements.

\begin{figure}
    \centering
    \includegraphics[width=1\textwidth]{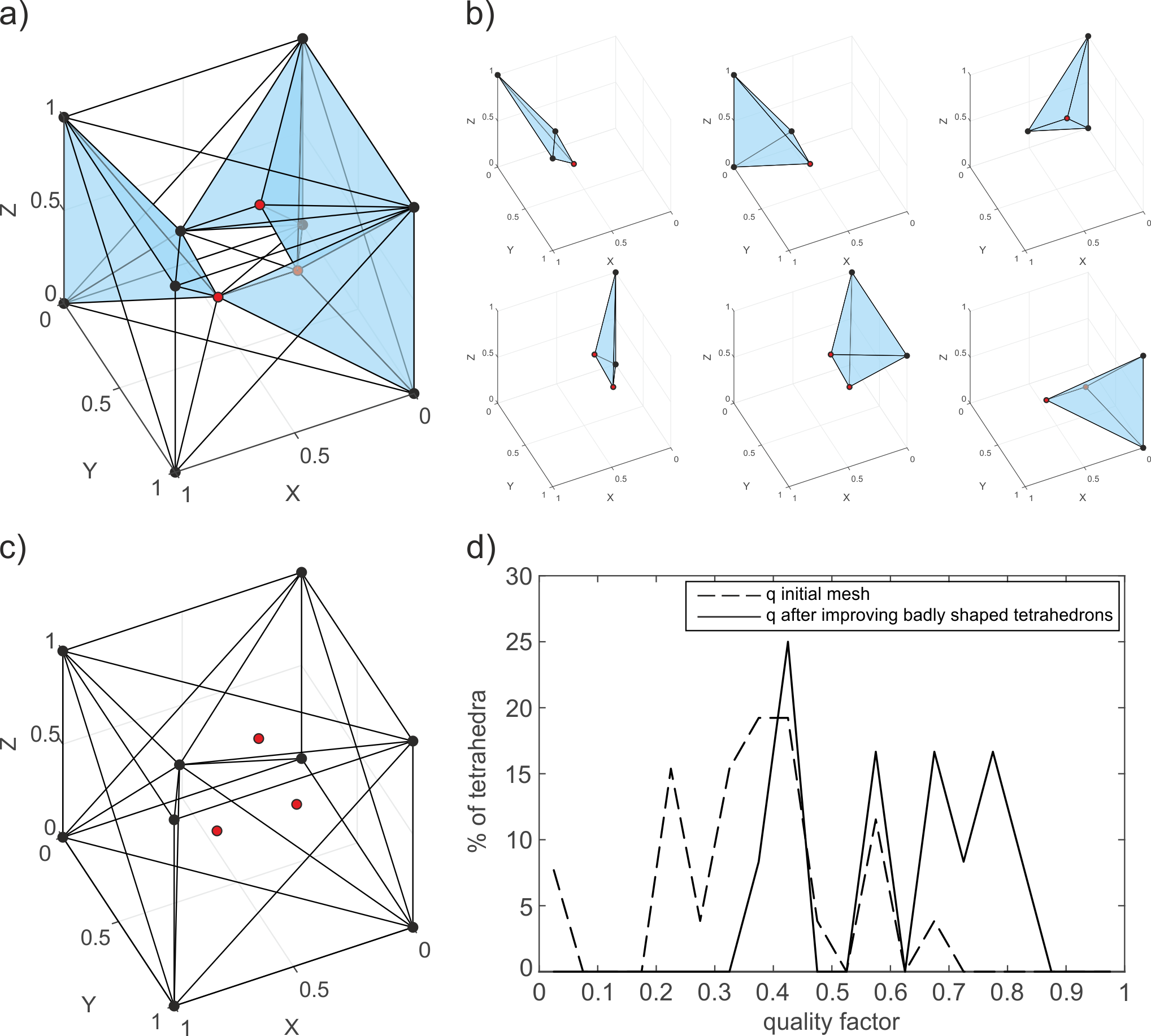}
    \caption{(a) Initial mesh with badly shaped tetrahedra (in blue). Rejected nodes in red. (b) Badly shaped tetrahedra. (c) Mesh after improving badly shaped tetrahedra contains no badly shaped tetrahedra. (d) Fraction of tetrahedra for a given quality factor for both before (dashed line) and after (solid line) local improvements to the shape of badly shaped tetrahedra. The minimum quality factor for the initial mesh is 0.04 and for the final mesh is 0.39.}
    \label{fig:improvement_bad_elements}
\end{figure}

\textit{\textbf{Improvement of badly shaped tetrahedra}}. Unstructured 3-D meshes are composed of irregular tetrahedra. Some may be quite poor in terms of their shape and quality factor (see \cite{Cheng2000} for a complete categorization of badly shaped tetrahedra). The first improvement for tetrahedral shapes acts locally and only modifies one node of each badly shaped tetrahedron. For each badly shaped tetrahedron, identified by $ q < q_{bad}$, where $ 0.2 \leq q_{bad} \leq 0.3 $, we select the spring with the maximum distortion, i.e. $ max(|\epsilon|) $. If $\epsilon > 0$, a new node is created in the midpoint of the selected spring, while a node at one end of the selected spring is removed if $\epsilon < 0$. A new connectivity is then created by another Delaunay triangulation. The new connectivity is only modified in the surroundings of nodes that have been added or removed, keeping the rest of the connectivity to be the same as the old triangulation. \cref{fig:improvement_bad_elements} illustrates a simple example that improves badly shaped tetrahedra when meshing the unit cube.

\begin{figure}
    \centering
    \includegraphics[width=1\textwidth]{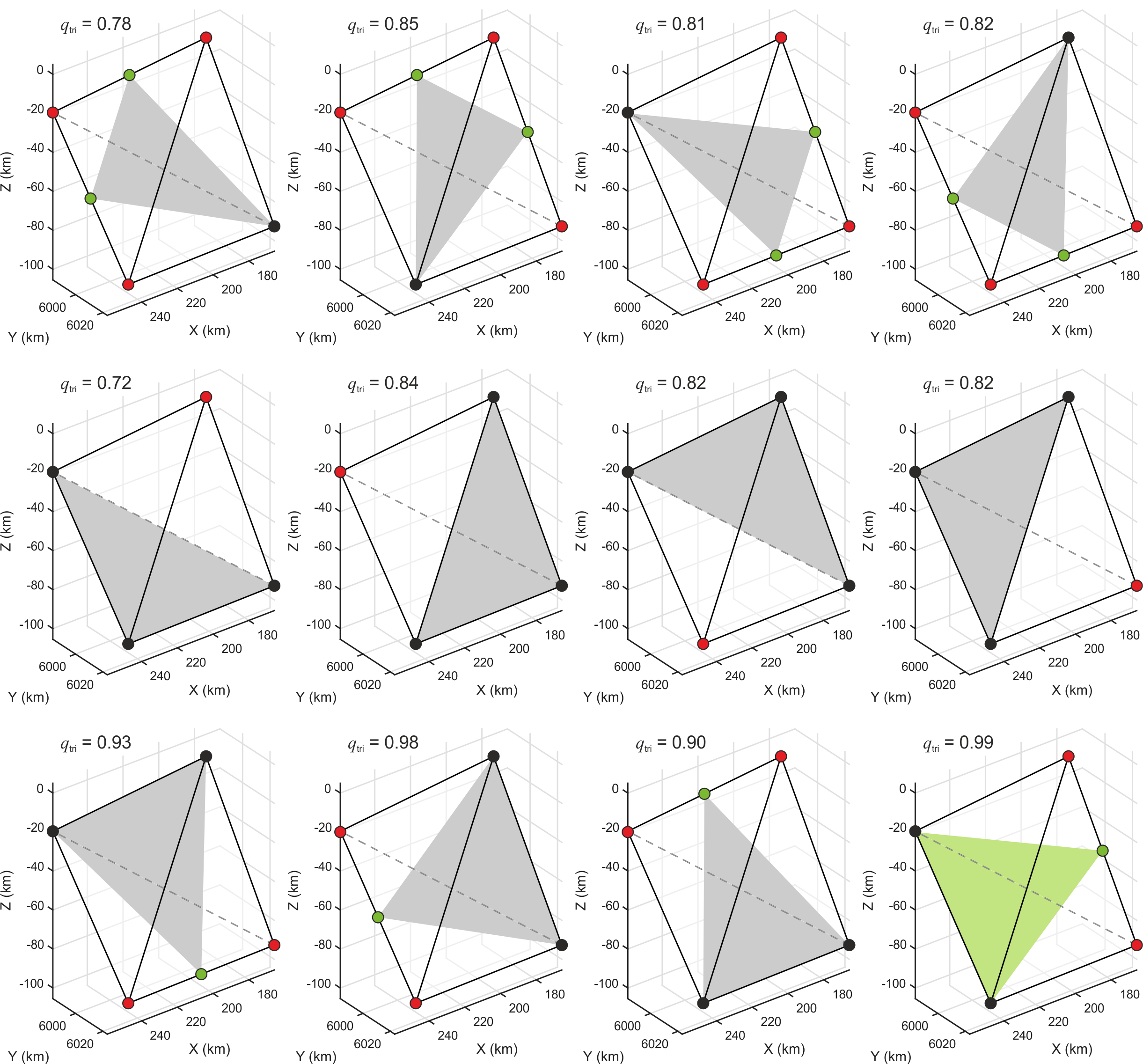}
    \caption{Removing a sliver (represented by black lines and dashed grey line for hidden edge). Possible triangles (grey and green colours) created from permutations of the vertices and midpoints of the edges of a sliver. Black, red and green points represent unaltered, removed and added nodes, respectively. $q_{tri}$ is the quality factor for each triangle. The four vertices of the sliver are replaced by the three mesh points of the potential triangle with the best quality factor (green colour).}
    \label{fig:slivers_permutation_v2}
\end{figure}

\textit{\textbf{Removing slivers}}. Slivers are degenerate tetrahedra whose vertices are well-spaced and near the equator of their circumsphere, hence their quality factor and enclosed volume are close to zero. We define a sliver as a tetrahedron with $ q < 0.1 $. Our routine for removing slivers is purely geometrical, i.e., it does not take into account the actual or desired length of the springs. The four vertices of each sliver are replaced by the three mesh points of the best potential triangle that can be generated from all permutations of its vertices and potential new nodes created at the midpoints of its springs (\cref{fig:slivers_permutation_v2}). Delaunay triangulation is called afterwards to create the connectivity matrix around the changed nodes.

\subsection*{Example: Spherical shell mesh with an embedded high resolution region}
\label{sec:Example: Spherical shell mesh with an embedded high resolution region}
We show the results for a spherical shell mesh with an embedded high-resolution sub-region (code available in \cref{SM5}). The input mesh parameters are listed in \cref{tab:1}. We recommend to set the point around which the refined region is created far from the polar axis since the guide-mesh can have difficulties in interpolating the desired spring lengths near the polar axis.

\begin{figure}
  	\centering
    \includegraphics[width=1\textwidth]{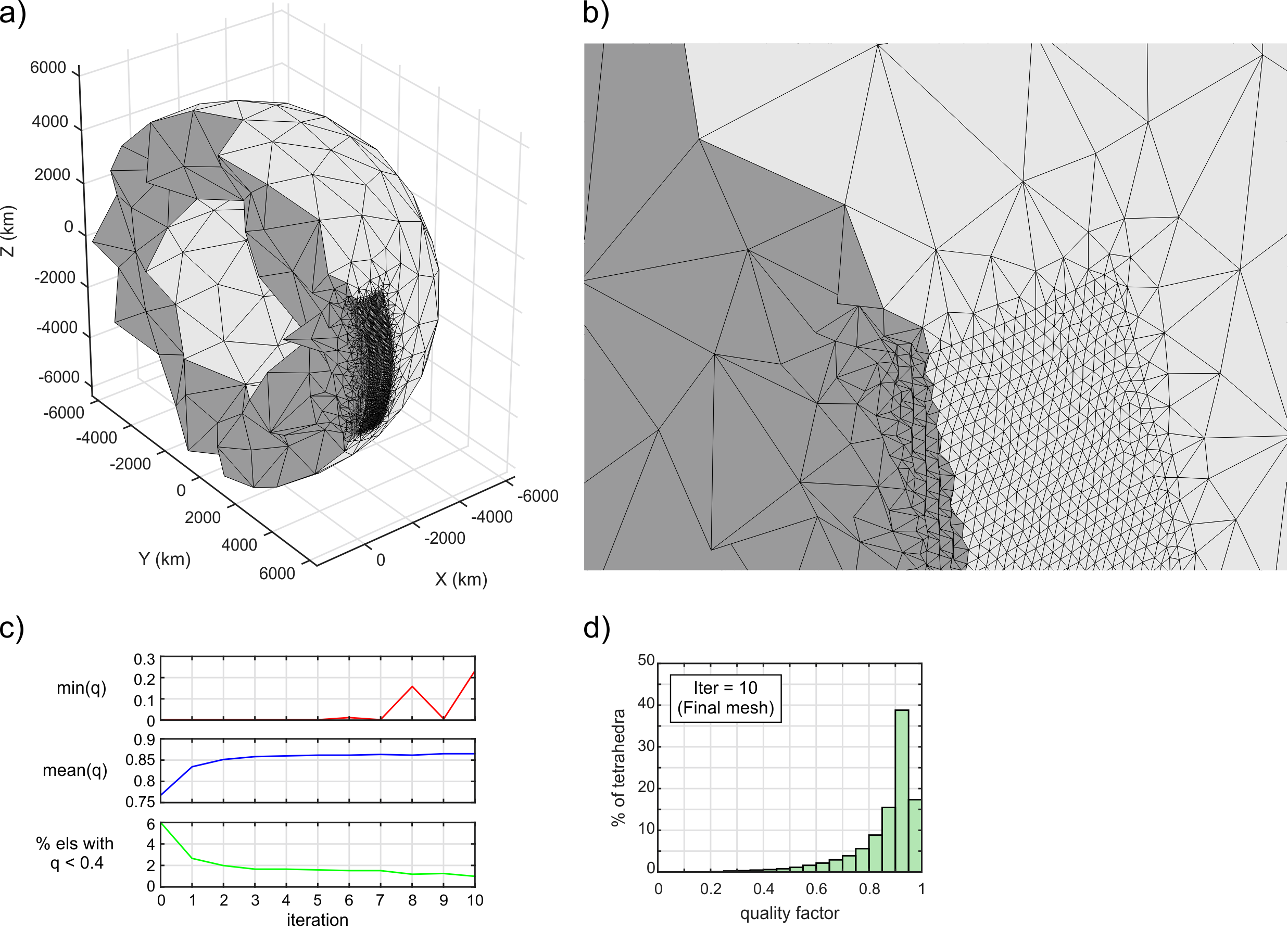}
  	\caption{(a) Cross section of the final mesh with an embedded high resolution sub-region after refinement using the guide-mesh. (b) Zoom around the boundary of the refined region. (c) Minimum quality factor (red line), mean quality factor for all elements (blue line) and fraction of elements having a quality factor lower than 0.4\% (green line) as a function of iteration number. (d) Histogram of the fraction of elements as a function of quality factor for the final mesh.}
  	\label{fig:results_SPH_v5}
\end{figure}

For this example, the domain of the mesh is a spherical shell whose boundaries represent the core-mantle boundary and the Earth's surface (Figure~\ref{fig:guide_mesh_SPH_ans_mesh_setup}b). The smallest tetrahedra with quasi-uniform size lie inside the high resolution region (red tesseroid in \cref{fig:guide_mesh_SPH_ans_mesh_setup}b). This region is embedded within a coarser global mesh. A transition region (green tesseroid in Figure~\ref{fig:guide_mesh_SPH_ans_mesh_setup}b) guarantees a gradual change in tetrahedral size from the high resolution region to the coarse region. The algorithm created the mesh in 224 s after 10 iterations (see \cref{fig:results_SPH_v5}a for a cross section of the final mesh). \cref{fig:results_SPH_v5}b shows a detail of the mesh around the northern boundary of the refined region. The mesh has 27000 nodes forming 150000 tetrahedra (\cref{tab:2}) with an edge-length factor $ l_{0r}/l_{0c} = 1/33$. The fraction of tetrahedra within the coarse, transition and refined regions is 0.8\%, 20.0\% and 79.2\% respectively (see \cref{fig:results_SPH_v5_SM}). The worst quality factor for an element is 0.23 (red line in \cref{fig:results_SPH_v5}c) and the mean of the quality factor for all elements is 0.87 (blue line in \cref{fig:results_SPH_v5}c). Only 1\% of the tetrahedra have a quality factor lower than 0.4. \cref{fig:results_SPH_v5}d shows the fraction of elements as a function of their quality factor for the final mesh.

\section{Summary}
\label{sec:Summary}
We have developed the tools for generating unstructured meshes in 2-D, 3-D, and spherical geometries that can contain embedded high resolution sub-regions. While we do not discuss the recipe for the (simpler) generation of a Cartesian 3-D mesh, only small modifications to the 3-D spherical code are needed to assign boundary points to lie along small sets of linear boundary edges and planar boundary surfaces. The algorithm employs the FEM to solve for the optimal nodal positions of a spring-like system of preferred nodal positions. Straight line, circular and spherical boundary conditions are imposed to constrain the shape of the mesh. We use a guide-mesh approach to smoothly refine the mesh around regions of interest. Methods for achieving the expected nodal density and improving the element shape and quality are also introduced to give robustness to the mesh. These allow it to make Finite Element meshes capable of higher computational accuracy and faster iterative convergence. This approach could also be extended to be used as part or all of an adaptive mesh refinement routine.

\section*{Acknowledgments}
We thank Cornell University for supporting the initial work on this problem by Morgan and Shi, and the COMPASS Consortium for the Ph.D. support for Jorge M. Taram\'on.

\bibliographystyle{siamplain}
\bibliography{main}

\newpage
\section*{SUPPLEMENTARY MATERIALS}

\subsection*{SM1. Derivation of equation \cref{eq:5}}
\label{SM1}
The 2-D development of equation \cref{eq:3}, rewritten here for convenience
\begin{equation}
  	\label{eq:A.1}
  	{\footnotesize
    \left(\begin{array}{c}
    {f_1}' \\
	{f_2}'
	\end{array}\right)
    + k
	\left[\begin{array}{rr}
		-1 & 1\\
		1 & -1
	\end{array}\right]
	\left(\begin{array}{c}
		0\\
		l_0
	\end{array}\right)
	= k
	\left[\begin{array}{rr}
		-1 &  1 \\
		 1 & -1
	\end{array}\right]
	\left(\begin{array}{c}
		{x_1}' \\
		{x_2}'
	\end{array}\right)
	}
\end{equation}
is given by two steps. First, develop the right hand side of equation \cref{eq:A.1} by writing local coordinates as a function of global coordinates (see \cref{fig:springs_2D_and_3D}a)
\begin{equation}
	\label{eq:A.2}
	{\footnotesize
	\begin{split}
	k
	\left[\begin{array}{rr}
	-1 &  1 \\
	 1 & -1
	\end{array}\right]
	\left(\begin{array}{c}
	{x_1}' \\
	{x_2}'
	\end{array}\right) \\
    &
	=
    k
	\left[\begin{array}{rr}
	{x_2}' - {x_1}' \\
	-({x_2}' - {x_1}')
	\end{array}\right] \\
    &
    = k
	\left[\begin{array}{rr}
	 \big[(x_2 - x_1)c_{\alpha} + (y_2 - y_1)s_{\alpha}\big] \\
	-\big[(x_2 - x_1)c_{\alpha} + (y_2 - y_1)s_{\alpha}\big]
	\end{array}\right] \\
	&
    = k
    \left[\begin{array}{rr}
	-1 &  1 \\
	 1 & -1
	\end{array}\right]
    \left[\begin{array}{rr}
	x_1 c_{\alpha} + y_1 c_{\alpha} \\
	x_2 c_{\alpha} + y_2 c_{\alpha}
	\end{array}\right] \\
	&
    =
    k
    \left[\begin{array}{rr}
	-1 &  1 \\
	 1 & -1
	\end{array}\right]
    \left[\begin{array}{cccc}
  	\cos\alpha & \sin\alpha &          0 &          0 \\
  	         0 &          0 & \cos\alpha & \sin\alpha
  	\end{array}\right]
    \left(\begin{array}{c}
	x_1 \\
	y_1 \\
	x_2 \\
    y_2
	\end{array}\right)
    \end{split}
    }
\end{equation}
where $s_\alpha\equiv\sin\alpha$ and $c_\alpha\equiv\cos\alpha$. Second, express the global coordinates of the force vector as a function of local coordinates (see \cref{fig:springs_2D_and_3D}a)
\begin{equation}
	\label{eq:A.3}
	{\footnotesize
	\left(\begin{array}{c}
	f_{1,x} \\
	f_{1,y} \\
	f_{2,x} \\
    f_{2,y} 
	\end{array}\right)
    =
    \left[\begin{array}{cc}
	c_{\alpha} & 0 \\
	s_{\alpha} & 0 \\
	0 		   & c_{\alpha}\\
    0 		   & s_{\alpha}
	\end{array}\right]
    \left(\begin{array}{c}
	{f_1}' \\
	{f_2}'
	\end{array}\right)
	}
\end{equation}
Combining equations \cref{eq:A.1,eq:A.2} gives
\begin{equation}
    \label{eq:A.4}
    {\footnotesize
    \left(\begin{array}{c}
    {f_1}' \\
	{f_2}'
	\end{array}\right)
	= k
    \left[\begin{array}{rr}
	-1 &  1 \\
	 1 & -1
	\end{array}\right]
    \left[\begin{array}{cccccc}
  	\cos\alpha & \sin\alpha &          0 &          0 \\
  	         0 &          0 & \cos\alpha & \sin\alpha
  	\end{array}\right]
    \left(\begin{array}{c}
	x_1 \\
	y_1 \\
	x_2 \\
    y_2
	\end{array}\right)
    - k
	\left[\begin{array}{rr}
		-1 & 1\\
		1 & -1
	\end{array}\right]
	\left(\begin{array}{c}
		0\\
		l_0
	\end{array}\right)
	}
\end{equation}
Substituting equation \cref{eq:A.4} into equation \cref{eq:A.3} and reordering gives
\begin{equation}
	{\footnotesize
    k
    \left[\begin{array}{cc}
	c_{\alpha} & 0 \\
	s_{\alpha} & 0 \\
	0 		   & c_{\alpha}\\
    0 		   & s_{\alpha}
	\end{array}\right]
    \left[\begin{array}{rr}
	-1 &  1 \\
	 1 & -1
	\end{array}\right]
    \left[\begin{array}{cccccc}
  	\cos\alpha & \sin\alpha &          0 &          0 \\
  	         0 &          0 & \cos\alpha & \sin\alpha
  	\end{array}\right]
    \left(\begin{array}{c}
	x_1 \\
	y_1 \\
	x_2 \\
    y_2
	\end{array}\right)
    \nonumber
    }
\end{equation}
\begin{equation}
	\label{eq:A.5}
	{\footnotesize
    =
    \left(\begin{array}{c}
	f_{1,x} \\
	f_{1,y} \\
	f_{2,x} \\
    f_{2,y}
	\end{array}\right)
    + k
    \left[\begin{array}{cc}
	c_{\alpha} & 0 \\
	s_{\alpha} & 0 \\
	0 		   & c_{\alpha}\\
    0 		   & s_{\alpha}
    \end{array}\right]
    \left[\begin{array}{rr}
	-1 &  1 \\
	 1 & -1
	\end{array}\right]
    \left(\begin{array}{c}
		0\\
		l_0
	\end{array}\right)
	}
\end{equation}
which is equivalent to equation \cref{eq:5}.

\subsection*{SM2. Derivation of equation \cref{eq:25}}
\label{SM2}
The 3-D development of equation \cref{eq:3}, rewritten here for convenience
\begin{equation}
  	\label{eq:B.1}
  	{\footnotesize
    \left(\begin{array}{c}
    {f_1}' \\
	{f_2}'
	\end{array}\right)
    + k
	\left[\begin{array}{rr}
		-1 & 1\\
		1 & -1
	\end{array}\right]
	\left(\begin{array}{c}
		0\\
		l_0
	\end{array}\right)
	= k
	\left[\begin{array}{rr}
		-1 &  1 \\
		 1 & -1
	\end{array}\right]
	\left(\begin{array}{c}
		{x_1}' \\
		{x_2}'
	\end{array}\right)
	}
\end{equation}
also involves two steps. First, develop the right hand side of equation (\ref{eq:B.1}) by writing local coordinates as a function of global coordinates (see \cref{fig:springs_2D_and_3D}b)
\begin{equation}
	\label{eq:B.2}
	{\footnotesize
	\begin{split}
	k
	\left[\begin{array}{rr}
	-1 &  1 \\
	 1 & -1
	\end{array}\right]
	\left(\begin{array}{c}
	{x_1}' \\
	{x_2}'
	\end{array}\right) \\
    &
	=
    k
	\left[\begin{array}{rr}
	{x_2}' - {x_1}' \\
	-({x_2}' - {x_1}')
	\end{array}\right] \\
    &
    = k
	\left[\begin{array}{rr}
	 \Big( \big[(x_2 - x_1)c_{\beta} + (y_2 - y_1)s_{\beta}\big]c_{\alpha} + (z_2 - z_1)s_{\alpha} \Big) \\
	-\Big( \big[(x_2 - x_1)c_{\beta} + (y_2 - y_1)s_{\beta}\big]c_{\alpha} + (z_2 - z_1)s_{\alpha} \Big)
	\end{array}\right] \\
	&
    = k
    \left[\begin{array}{rr}
	-1 &  1 \\
	 1 & -1
	\end{array}\right]
    \left[\begin{array}{rr}
	x_1c_{\alpha}c_{\beta} + y_1c_{\alpha}s_{\beta}  + z_1s_{\alpha} \\
	x_2c_{\alpha}c_{\beta}  + y_2c_{\alpha}s_{\beta}  + z_2s_{\alpha}
	\end{array}\right] \\
	&
    =
    k
    \left[\begin{array}{rr}
	-1 &  1 \\
	 1 & -1
	\end{array}\right]
    \left[
	\arraycolsep=1.4pt    
    \begin{array}{cccccc}
  	c_{\alpha}c_{\beta} & c_{\alpha}s_{\beta} & s_{\alpha} &                   0 &                   0 &          0 \\
  	                  0 &                   0 &          0 & c_{\alpha}c_{\beta} & c_{\alpha}s_{\beta} & s_{\alpha}
  	\end{array}\right]
    \left(\begin{array}{c}
	x_1 \\
	y_1 \\
    z_1 \\
	x_2 \\
    y_2 \\
	z_2
	\end{array}\right)
    \end{split}
    }
\end{equation}
where $s_\alpha\equiv\sin\alpha$, $c_\alpha\equiv\cos\alpha$, $s_\beta\equiv\sin\beta$ and $c_\beta\equiv\cos\beta$. Second, express the global coordinates of the force vector as a function of local coordinates (see \cref{fig:springs_2D_and_3D}b)
\begin{equation}
	\label{eq:B.3}
	{\footnotesize
	\left(\begin{array}{c}
	f_{1,x} \\
	f_{1,y} \\
    f_{1,z} \\
	f_{2,x} \\
    f_{2,y} \\
	f_{2,z}
	\end{array}\right)
    =
    \left[\begin{array}{cc}
	c_{\alpha}c_{\beta} & 0 \\
	c_{\alpha}s_{\beta} & 0\\
    s_{\alpha} 			& 0\\
	0 					& c_{\alpha}c_{\beta}\\
    0 					& c_{\alpha}s_{\beta}\\
	0					& s_{\alpha}
	\end{array}\right]
    \left(\begin{array}{c}
	{f_1}' \\
	{f_2}'
	\end{array}\right)
	}
\end{equation}
Combining equations \cref{eq:B.1,eq:B.2} gives
\begin{equation}
    \label{eq:B.4}
    {\footnotesize
    \left(\begin{array}{c}
    {f_1}' \\
	{f_2}'
	\end{array}\right)
	= k
    \left[\begin{array}{rr}
	-1 &  1 \\
	 1 & -1
	\end{array}\right]
    \left[
	\arraycolsep=1.4pt    
    \begin{array}{cccccc}
  	c_{\alpha}c_{\beta} & c_{\alpha}s_{\beta} & s_{\alpha} &                   0 &                   0 &          0 \\
  	                  0 &                   0 &          0 & c_{\alpha}c_{\beta} & c_{\alpha}s_{\beta} & s_{\alpha}
  	\end{array}\right]
    \left(\begin{array}{c}
	x_1 \\
	y_1 \\
    z_1 \\
	x_2 \\
    y_2 \\
	z_2
	\end{array}\right)
    - k
	\left[\begin{array}{rr}
		-1 & 1\\
		1 & -1
	\end{array}\right]
	\left(\begin{array}{c}
		0\\
		l_0
	\end{array}\right)
	}
\end{equation}
Substituting equation \cref{eq:B.4} into equation \cref{eq:B.3} and reordering gives
\begin{equation}
	{\footnotesize
    k
    \left[\begin{array}{cc}
	c_{\alpha}c_{\beta} & 0 \\
	c_{\alpha}s_{\beta} & 0\\
    s_{\alpha} 			& 0\\
	0 					& c_{\alpha}c_{\beta}\\
    0 					& c_{\alpha}s_{\beta}\\
	0					& s_{\alpha}
	\end{array}\right]
    \left[\begin{array}{rr}
	-1 &  1 \\
	 1 & -1
	\end{array}\right]
    \left[
	\arraycolsep=1.4pt    
    \begin{array}{cccccc}
  	c_{\alpha}c_{\beta} & c_{\alpha}s_{\beta} & s_{\alpha} &                   0 &                   0 &          0 \\
  	                  0 &                   0 &          0 & c_{\alpha}c_{\beta} & c_{\alpha}s_{\beta} & s_{\alpha}
  	\end{array}\right]
    \left(\begin{array}{c}
	x_1 \\
	y_1 \\
    z_1 \\
	x_2 \\
    y_2 \\
	z_2
	\end{array}\right)
    \nonumber
    }
\end{equation}
\begin{equation}
	\label{eq:B.5}
	{\footnotesize
    =
    \left(\begin{array}{c}
	f_{1,x} \\
	f_{1,y} \\
    f_{1,z} \\
	f_{2,x} \\
    f_{2,y} \\
	f_{2,z}
	\end{array}\right)
    + k
    \left[\begin{array}{cc}
	c_{\alpha}c_{\beta} & 0 \\
	c_{\alpha}s_{\beta} & 0\\
    s_{\alpha} 			& 0\\
	0 					& c_{\alpha}c_{\beta}\\
    0 					& c_{\alpha}s_{\beta}\\
	0					& s_{\alpha}
    \end{array}\right]
    \left[\begin{array}{rr}
	-1 &  1 \\
	 1 & -1
	\end{array}\right]
    \left(\begin{array}{c}
		0\\
		l_0
	\end{array}\right)
	}
\end{equation}
which is equivalent to equation \cref{eq:25}.

\subsection*{SM3. Code for Rectangular mesh generation}
\label{SM3}
\begin{figure}
  	\centering
    \includegraphics[width=1\textwidth]{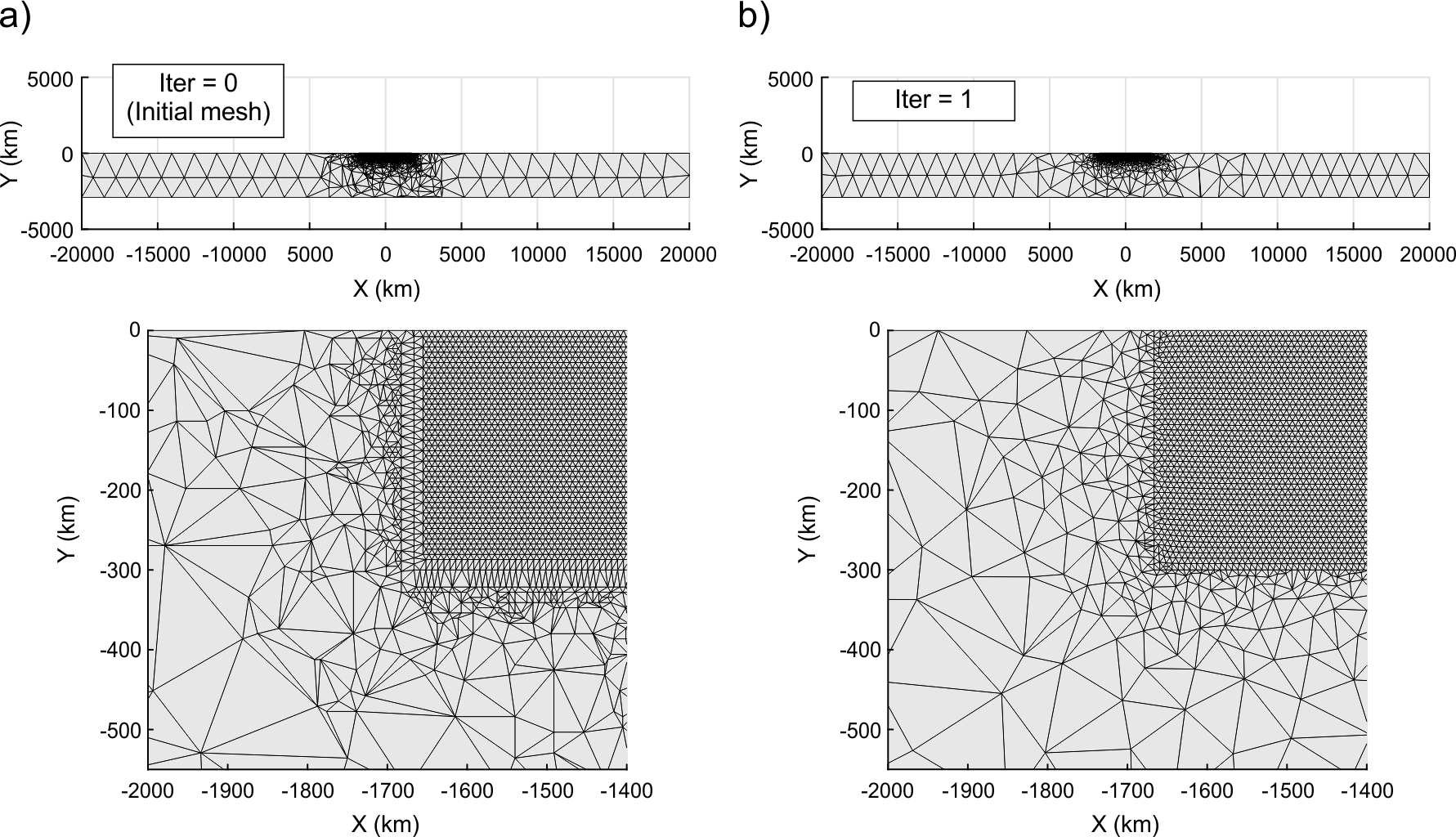}
  	\caption{(a) Initial mesh (top) for a rectangular box with an embedded high resolution sub-region and a zoom around the left boundary of the refined region (bottom). (b) Mesh (top) and zoom (bottom) after the first iteration.}
  	\label{fig:results_RECT_v4_SM}
\end{figure}

Code to reproduce the example shown in \cref{fig:results_RECT_v4} and \cref{fig:results_RECT_v4_SM}

\subsection*{SM4. Code for Cylindrical annulus mesh generation}
\label{SM4}
\begin{figure}
  	\centering
    \includegraphics[width=1\textwidth]{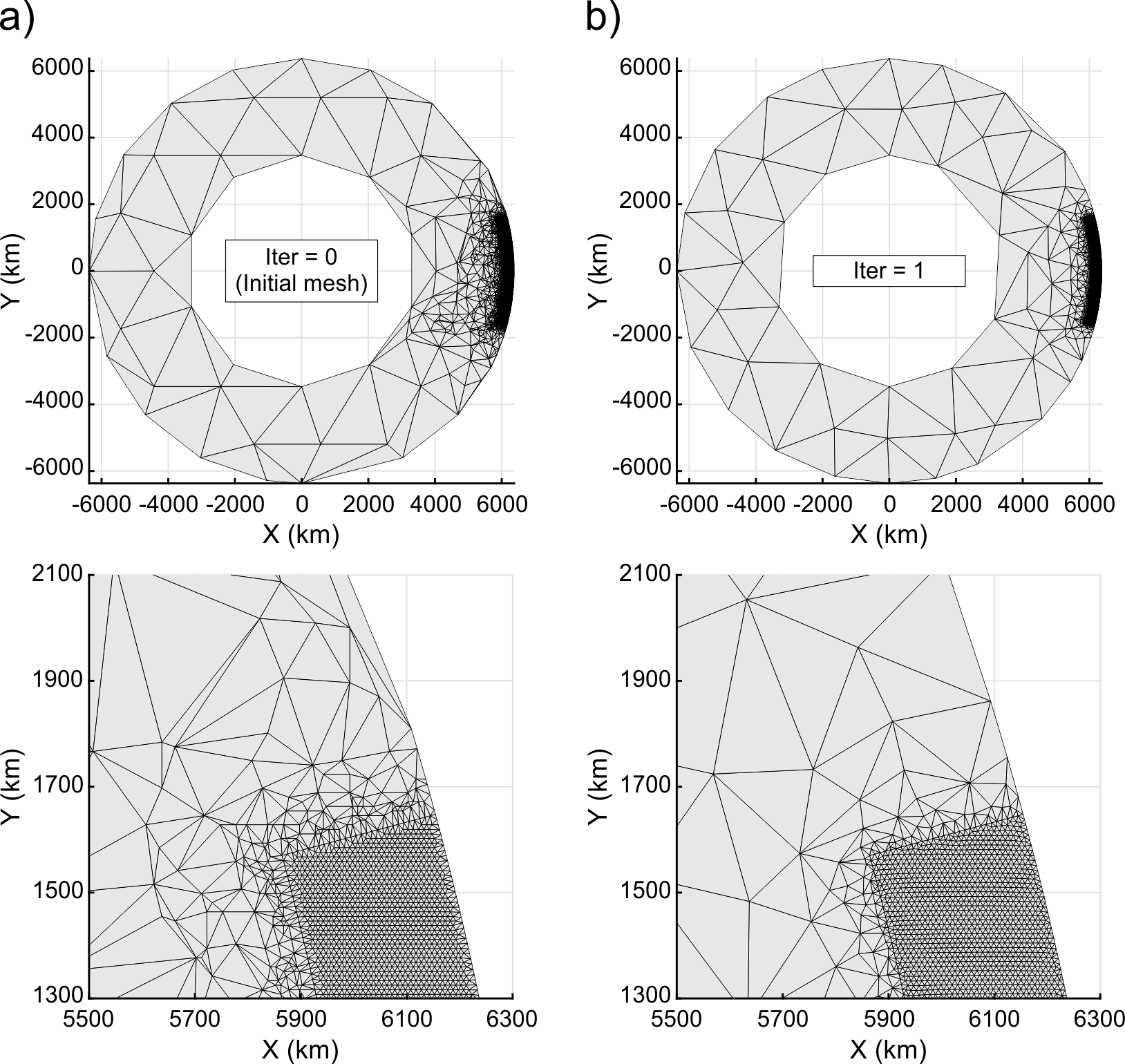}
  	\caption{(a) Initial mesh (top) for a cylindrical annulus with an embedded high resolution sub-region and a zoom around an edge of the refined region (bottom). (b) Mesh (top) and zoom (bottom) after the first iteration.}
  	\label{fig:results_CYL_v4_SM}
\end{figure}

Code to reproduce the example shown in \cref{fig:results_CYL_v4} and \cref{fig:results_CYL_v4_SM}

\subsection*{SM5. Code for Spherical shell mesh generation}
\label{SM5}
\begin{figure}
  	\centering
    \includegraphics[width=1\textwidth]{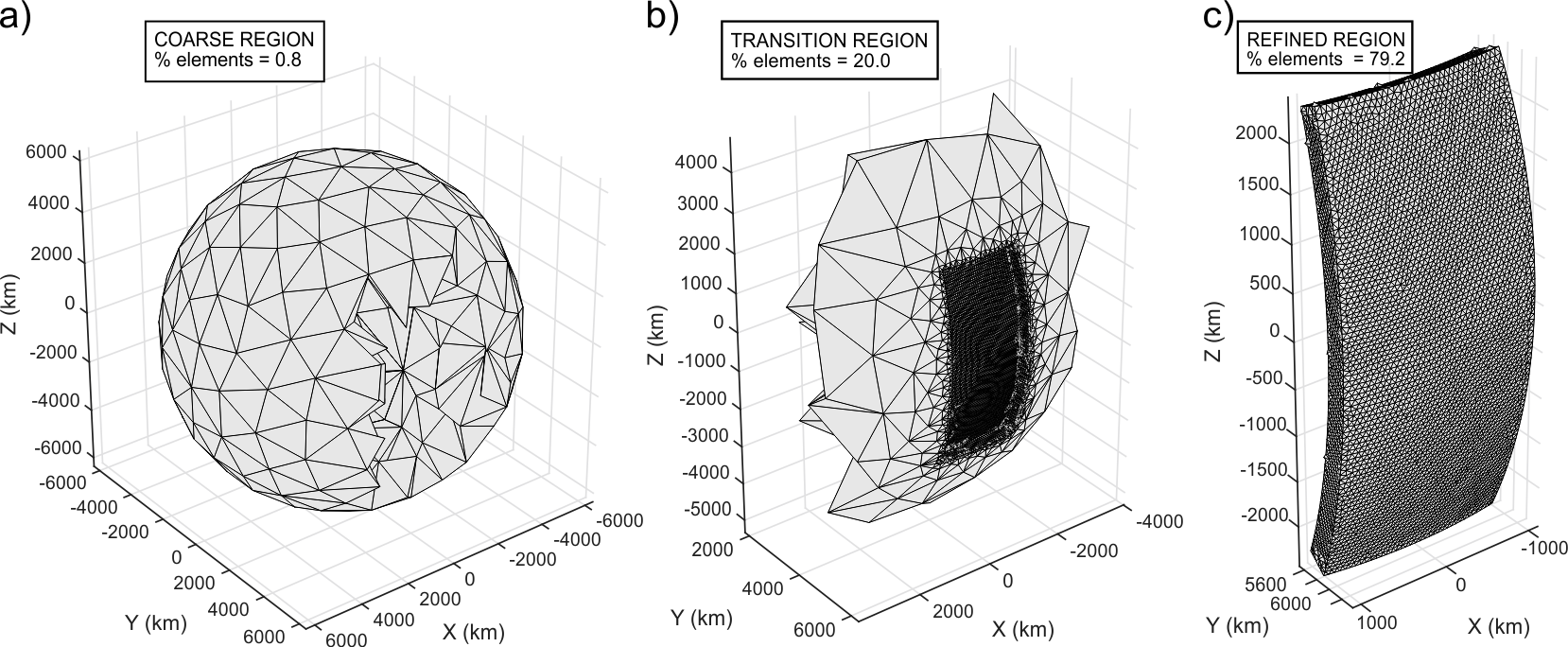}
  	\caption{(a) Tetrahedra within the coarse region. (b) Tetrahedra within the transition region. (c) Tetrahedra within the refined region.}
  	\label{fig:results_SPH_v5_SM}
\end{figure}

Code to reproduce the example shown in \cref{fig:results_SPH_v5} and \cref{fig:results_SPH_v5_SM}

\subsection*{SM6. Additional Figures}
\label{SM6}

\end{document}

%% file: shared.tex
\usepackage{lipsum}
\usepackage{amsfonts}
\usepackage{graphicx}
\usepackage{epstopdf}
\usepackage{algorithmic}
\ifpdf
  \DeclareGraphicsExtensions{.eps,.pdf,.png,.jpg}
\else
  \DeclareGraphicsExtensions{.eps}
\fi

\numberwithin{theorem}{section}

\newcommand{\TheTitle}{Generation of unstructured meshes in 2-D, 3-D, and spherical geometries with embedded high resolution sub-regions} 
\newcommand{\TheAuthors}{J. M. Taram\'on, J. P. Morgan, C. Shi, and J. Hasenclever}

\headers{Mesh generation with embedded hi-res sub-regions}{\TheAuthors}

\title{{\TheTitle}\thanks{Submitted to the editors 14Nov2017.
\funding{This work was funded by the COMPASS Consortium.}}}

\author{
  Jorge M. Taram\'on\thanks{Department of Earth Sciences, Royal Holloway University of London, Egham, UK
    (\email{Jorge.TaramonGomez.2014@live.rhul.ac.uk}).}
  \and
  Jason P. Morgan\footnotemark[2] \thanks{EAS Dept., Snee Hall, Cornell University, Ithaca, NY, USA}
  \and
  Chao Shi\footnotemark[3]
  \and
  J\"org Hasenclever\thanks{Institute of Geophysics, Hamburg University, Hamburg, Germany}
}

\usepackage{amsopn}
